\documentclass[fleqn,usenatbib]{mnras}

\usepackage[T1]{fontenc}
\usepackage{ae,aecompl}

\usepackage{graphicx}	
\usepackage{amsmath}	
\usepackage{amssymb}	
\usepackage{subfigure}

\title[The dynamics of NGC~3201]{The dynamics of the globular cluster NGC~3201 out to the Jacobi radius}

\author[Wan et al.]{Zhen Wan$^{1}$\thanks{E-mail: zhen.wan@sydney.edu.au},
William H. Oliver$^{1}$,
Holger Baumgardt$^{2}$,
Geraint F. Lewis$^{1}$, \newauthor
Mark Gieles$^{3,4}$,
Vincent H\'{e}nault-Brunet$^{5}$,
Thomas de Boer$^{6,7}$,
Eduardo Balbinot$^{8}$, \newauthor
Gary Da Costa$^{9}$,
and Dougal Mackey$^{9}$
\\ \\
$^{1}$Sydney Institute for Astronomy, School of Physics A28, The University of Sydney, NSW, 2006, Australia\\
$^{2}$School of Mathematics and Physics, The University of Queensland, St Lucia, QLD 4072, Australia\\
$^{3}$ICREA, Pg. Llu\'{i}s Companys 23, E08010 Barcelona, Spain\\
$^{4}$Institut de Ci\`{e}ncies del Cosmos (ICCUB), Universitat de Barcelona (IEEC-UB), Mart\'{i} Franqu\`{e}s 1, E08028 Barcelona, Spain \\
$^{5}$Department of Astronomy and Physics, Saint Mary's University, 923 Robie Street, Halifax, NS B3H 3C3, Canada\\
$^{6}$Institute for Astronomy, University of Hawaii, 2680 Woodlawn Drive, Honolulu, HI 96822, USA\\
$^{7}$Department of Physics, University of Surrey, Guildford GU2 7XH, UK\\
$^{8}$Kapteyn Astronomical Institute, University of Groningen, Postbus 800, NL-9700AV Groningen, The Netherlands\\
$^{9}$Research School of Astronomy and Astrophysics, Australian National University, Canberra, ACT 2611, Australia\\
}

\date{Accepted XXX. Received YYY; in original form ZZZ}

\pubyear{2020}

\usepackage{newtxtext,newtxmath}
\begin{document}
\label{firstpage}
\pagerange{\pageref{firstpage}--\pageref{lastpage}}
\maketitle

\begin{abstract}
As part of a chemo-dynamical survey of five nearby globular clusters with 2dF/AAOmega on the Anglo-Australian Telescope (AAT), we have obtained kinematic information for the globular cluster NGC~3201. 
Our new observations confirm the presence of a significant velocity gradient across the cluster which can almost entirely be explained by the high proper motion of the cluster ($\sim9\,\mathrm{mas\,yr^{-1}}$).
After subtracting the contribution of this perspective rotation, we found a remaining rotation signal with an amplitude of $\sim1\,\mathrm{km\,s^{-1}}$ around a different axis to what we expect from the tidal tails and the potential escapers, suggesting that this rotation is internal and can be a remnant of its formation process. At the outer part, we found a rotational signal that is likely a result from potential escapers. The proper motion dispersion at large radii reported by Bianchini et al. ($3.5\pm0.9\,\mathrm{km\,s^{-1}}$) has previously been attributed to dark matter. Here we show that the LOS dispersion between $0.5-1$ Jacobi radius is lower ($2.01\pm0.18\,\mathrm{km\,s^{-1}}$), yet above the predictions from an $N$-body model of NGC~3201 that we ran for this study ($1.48\pm0.14\,\mathrm{km\,s^{-1}}$). Based on the simulation, we find that potential escapers cannot fully explain the observed velocity dispersion. We also estimate the effect on the velocity dispersion of different amounts of stellar-mass black holes and unbound stars from the tidal tails with varying escape rates and find that these effects cannot explain the difference between the LOS dispersion and the $N$-body model. Given the recent discovery of tidal tail stars at large distances from the cluster, a dark matter halo is an unlikely explanation.
We show that the effect of binary stars, which is not included in the $N$-body model, is important and can explain part of the difference in dispersion. We speculate that the remaining difference must be the result of effects not included in the $N$-body model, such as initial cluster rotation, velocity anisotropy and Galactic substructure.

\end{abstract}



                                                                                 
\section{Introduction}

The formation and evolution  of globular clusters (GC) remains an open question in astrophysics. 
Clues from dynamical signatures have proven to be useful to unravelling this, with structures in GC phase-space, such as tidal arms and velocity gradients along the length of these arms \citep[e.g.][]{2010A&A...522A..71J,2010AJ....139..606C,2011ApJ...726...47S,2020arXiv200712165H} representing evidence of the interaction between GCs and their host galaxies. Similarly, internal dynamical features including rotation \citep[e.g.][]{2019MNRAS.485.1460S,2019MNRAS.489..623V,2012A&A...538A..18B,2018MNRAS.473.5591K,2018ApJ...860...50F,2018A&A...616A..12G,2018MNRAS.481.2125B,2018ApJ...861...16L,2018ApJ...865...11L}, stellar envelopes \citep{2014MNRAS.442.3044M,2018MNRAS.473.2881K,2019MNRAS.485.4906D}, and the velocity dispersion profile \citep[e.g.][]{2003A&A...405L..15S,2007A&A...462L...9S,2010MNRAS.407.2241K,2018MNRAS.478.1520B,baumgardtetal2019} are thought to trace both the formation and evolution of GCs. 

However, in answering questions surrounding whether GCs are born \textit{in-situ} or \textit{ex-situ}, key information such as a GC's formation environment, or the time taken for a GC to be accreted into its galactic host, still remain unclear. In particular, whether or not GCs are born within dark matter mini-halos is still under debate, although given their extreme age, theoretical models have suggested that GC formation occurs 
within a dark matter mini-halo of a mass of $\sim 10^{8}\,\mathrm{M_{\odot}}$  \citep{1984ApJ...277..470P,2015ApJ...808L..35T}. 
The presence of stellar envelopes surrounding some GCs -- where stars are confined to the GC over a long time period -- is in agreement with this theory \citep{2012MNRAS.419...14C,2017MNRAS.471L..31P,2018MNRAS.473.2881K}, whereas the presence of tidal features \citep[e.g.][]{2001ApJ...548L.165O} is not \citep{1996ApJ...461L..13M}, and the absence of tidal tails in some GCs can in some cases be explained by the preferential loss of low-mass stars due to mass segregation \citep{2018MNRAS.474.2479B}.
Furthermore, whilst the comparison between the dynamics and stellar luminosity in the inner parts of GCs suggests that it is not necessary to include non-baryonic dark matter \citep[e.g.][]{2011ApJ...741...72C, 2015ApJ...803...29W,2015AJ....149...53K,2017MNRAS.464.2174B,2018MNRAS.473.4832G}, this is not evidence of the absence of dark matter in the outer regions of GCs; collisional relaxation can push the dark matter to the periphery where tidal interaction with the Milky Way (MW) is effective in stripping the entire dark matter content \citep{2005ApJ...619..243M,2005ApJ...619..258M,2008MNRAS.391..942B}. 

The dynamics at the periphery of GCs is where different models can be distinguished since the presence of dark matter inevitably inflates the velocity dispersion here. The existence of `potential escapers' within this region creates some difficulty when testing these models. These stars are located within the Jacobi radius, but have energies above the critical energy for escape  \citep{2000MNRAS.318..753F,2001MNRAS.325.1323B,2017MNRAS.466.3937C,2017MNRAS.468.1453D}. As a result, the outer density profiles of GCs can be very similar to the dark matter prediction \citep{2010MNRAS.407.2241K}, though finding the retrograde rotation of potential escapers \citep{2016MNRAS.461..402T} would strongly support the scenario where GCs do not possess dark matter, at least not at present.

Direct imaging of stars in many GCs out to large radii is available \citep[e.g.][]{2018MNRAS.476..271S}, although spectroscopic observations are still lacking for many stars beyond half the Jacobi radius \citep{2017MNRAS.466.3937C}. This results from target selection based solely on colour-magnitude diagrams (CMDs) where MW stars significantly outnumber the cluster members in the low-density outskirts of GCs, resulting in a low efficiency when allocating spectroscopic fibres. However, this changed with the arrival of \textit{Gaia} Data Release 2 \citep{2018A&A...616A...1G,2018A&A...616A..10G} which includes the precise proper motion measurements of distant halo stars, allowing the isolation of GC members using both their photometry and their astrometry. Using this catalogue, we have selected a sample of GC members and performed a spectroscopic survey of five nearby GCs -- NGC~3201, NGC~1904, NGC~1851, NGC~1261 and NGC~4590 -- with 2dF/AAOmega on the 3.9-m Anglo-Australian Telescope. We direct our efforts towards stars situated beyond half the Jacobi radius with the aim of understanding the dynamics of these GCs with the resulting moderate resolution spectra of their members.

\begin{figure}
	\centering
	\includegraphics[width=\columnwidth]{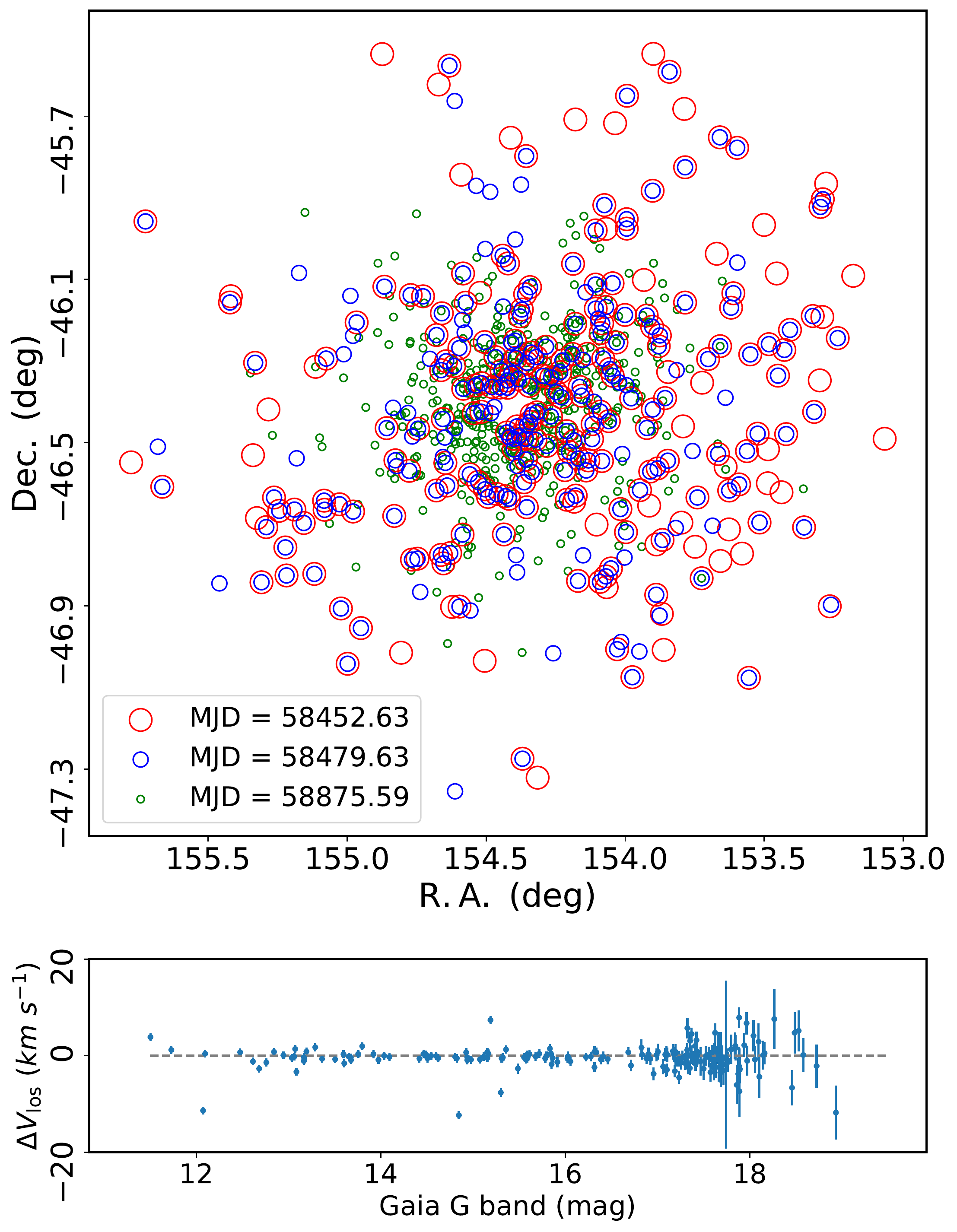}
	\caption{The top panel shows the footprint of our survey of NGC~3201, where different epochs are marked in different coloured and sized circles; some stars have multi-epoch observations so that we can analyse the impact of binaries. The bottom panel presents the velocity difference between the first and second epochs, where some binaries deviate significantly from the zero line.}
	\label{fig:N3201_footprint}	
\end{figure}

In this paper, we present a brief summary of our survey, and with the longest exposure time, we present the first scientific results on NGC~3201. This cluster is an interesting GC with its retrograde orbit, which is assigned as accreted in the Gaia-Enceladus/Sequoia event by \citet[][]{2019A&A...630L...4M}. We discuss the details of the survey in Sec.~\ref{sec:obs}, including the target selection, observations and data reduction. To interpret the observations of NGC~3201, we compare our data to an $N$-body simulation in Sec.~\ref{sec:nbody}. We discuss our first results on NGC~3201 as well as the effects from binary stars and black holes (BHs) in Sec.~\ref{sec:results} and present our conclusions in Sec.~\ref{sec:conclusions}. We note that results on the remainder of the survey GCs will be published in later contributions.
\begin{figure*}
    \centering
    \includegraphics[width=2\columnwidth]{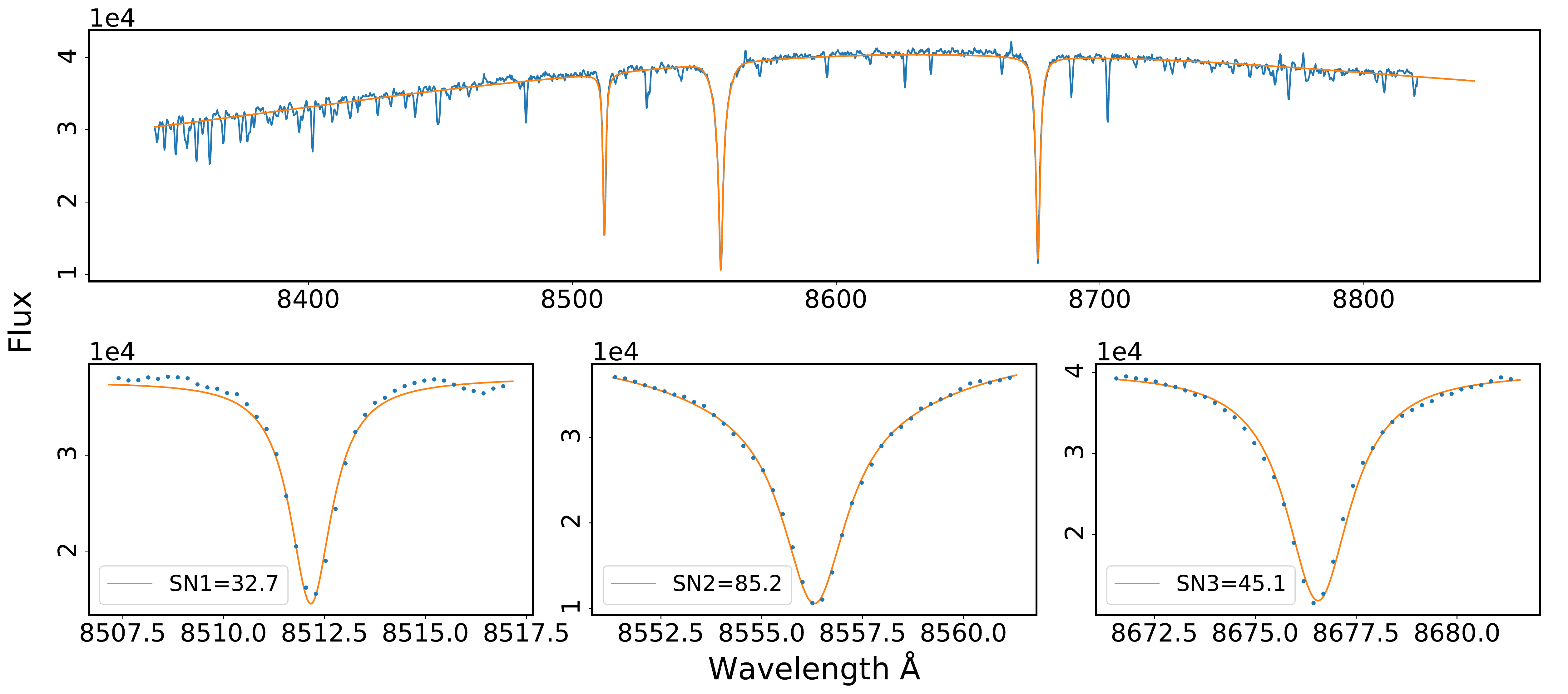}
    \caption{An example of a spectrum with a high S/N. The top panel shows the spectra in blue as well as the best-fitting CaT profile in orange. The bottom three panels are zoomed-in views of the three absorption lines.}
    \label{fig:sample}
\end{figure*}

\section{Observations and data reduction}
\label{sec:obs}
\subsection{Target selection}

The selection of targets for spectroscopic follow-up is based on the samples of GC members using data from {\it Gaia} DR2 produced by \citet{2019MNRAS.485.4906D}. The GC samples are extracted through the application of a `matched-filter' algorithm to the colour-magnitude diagrams, using an isochrone from the Padova library \citep{Marigo17}, as queried from \url{http://stev.oapd.inaf.it/cmd}. The GC metallicity ($\mathrm{[Fe/H]} = -1.59$) and distance (4.9 kpc) are taken from \citet{Harris10}, with the age (11.5 Gyr) taken from~\citet{Marin-Franch09,VandenBerg13}. For a more secure membership selection, we consider only stars in a region around the isochrone with $\vert(G_{\rm BP}-G_{\rm RP}) - (G_{\rm BP}-G_{\rm RP})_{0}\vert < 2\times\delta(G_{\rm BP}-G_{\rm RP})$ at each $G$ magnitude, with a minimum colour error of 0.03. Here the $G_{\rm BP}$ and $G_{\rm RP}$ represent the magnitude in the {\it Gaia} GP and RP bands, and the $\delta(G_{\rm BP}-G_{\rm RP})$ is the colour error.

The sample is further cleaned using {\it Gaia} DR2 proper motions to compute the membership probability of each star. The proper motions are fit using a Gaussian mixture model consisting of a cluster distribution and a MW foreground distribution. Initial guesses for the cluster Gaussian centres are taken from~\citet{Helmi18}, before distributions are fit using the $\texttt{emcee}$ python MCMC package~\citep{2013PASP..125..306F}. The parameters of the symmetric 2D Gaussian for NGC~3201 are [$\mu_{\mathrm{ra}}$, $\mu_{\mathrm{dec}}$, $\sigma$] = [8.37$\pm$0.12, -1.96$\pm$0.12,  0.47$\pm$0.15] $\mathrm{mas\,yr^{-1}}$. The final member samples are then selected by adopting a cut of 0.5 for the proper motion membership probability. To assess the importance of the various selection cuts, we note that the initial sample of 623583 stars is reduced to 79480 following the colour cuts and reduced further to 9913 given the proper motion selection. Therefore, applying the {\it Gaia} DR2 proper motion cuts is instrumental in obtaining a robust sample of high probability members that can reasonably be followed up with spectroscopic facilities.

The resulting samples of members cover the entire spatial extent of the GCs we intend to study, with proper motion errors of $0.6\,\mathrm{mas\,yr^{-1}}$ at {\it Gaia} $G=19\,$mag. For this survey, we focus on the Ca \textsc{ii} triplet (CaT) at 8498.02 \hbox{\AA}, 8542.09 \hbox{\AA} and 8662.14 \hbox{\AA} \citep{edlen1956spectrum}, hence all targets were selected from the red giant branch in each GC.

For our sample of NGC~3201 members we find that there are $\approx$10000 member stars available within 2dF's 2-degree field of view with 1944 members beyond 0.25$r_{\rm Jacobi}$. This ensures that enough fibres can be allocated outside of the densely crowded central regions of the GCs. The radii probed by 2dF are well outside the range of currently available data (within 15 arcmins from the GC centre) and contain sufficient numbers of cluster members to measure a possible bulk cluster rotation that is retrograde with respect to the orbit of the GC (due to potential escapers).

\subsection{Observations with AAT and 2dF/AAOmega}
The AAT is a 3.9m optical telescope located at Siding Spring Observatory near Coonabarabran, New South Wales, Australia. For this study, we used 2dF/AAOmega, which is a fibre positioner with a field of view of 2 degrees coupled to a dual-arm spectrograph. We use the 580V grating for the blue arm and the 1700D grating for the red arm, corresponding to resolutions of $\sim$ 1300 and $\sim$ 10000, and wavelength ranges of 3800-5800 \hbox{\AA} and 8400-8820 \hbox{\AA} respectively. 
The red arm setting enables the coverage of the target CaT lines, and, given the pixel resolution, results in velocity uncertainties of $\sim 1\,\mathrm{km\,s^{-1}}$ with a signal-to-noise ratio of 10. 

We use \textsc{configure} \citep{2006MNRAS.371.1537M} to produce the configuration file for 2dF for each target list, and acquire sufficient biases and calibrations for the data reduction during observing nights. The total integration time per field pointing was 2 hours, split into four individual exposures of 30 minutes to mitigate the effects of cosmic rays.

When possible, to mitigate the effects of binaries, we split the observations into multiple epochs with a separation of about a month or more. More specifically, observations were performed in three blocks; 29-30 Nov 2018, 27-30 Dec 2018, and 27-28 Jan 2020. Overall we obtained $\sim$ 4 nights-worth of useful observing time, during which the seeing ranged from $1.7$--$\sim3\,\mathrm{arcsec}$. We obtained a single epoch for each of NGC~1261 and NGC~4590, and multiple epochs for NGC~3201, NGC~1904 and NGC~1851. The observational epochs and exposure information are summarised in Table~\ref{tab:observation_log}, and the survey footprint across NGC~3201 is presented in Fig.~\ref{fig:N3201_footprint}.

\begin{table*}
    \centering
    \begin{tabular}{c|c|c|c|c|c}
        \hline
        Target Name & Epochs & exposure time & $\mathrm{N_{target}}$ & $\mathrm{N_{good star}}$ & MJD  \\
        \hline
        NGC 3201 & 3 & 7200s/ 7200s/ 9000s + 3600s & 248/ 252/ 321 + 147 & 207/ 213/ 320 + 146 & 58452.63/ 58479.63/ 58875.59\\
        NGC 1904 & 2 & 7800s/ 7200s & 188/ 77 & 121/ 45 & 58452.53/ 58875.49 \\
        NGC 1851 & 2 & 7200s/ 1800s & 126/74 & 95/58 & 58479.54/ 58876.53\\
        NGC 1261 & 1 & 7800s & 138 & 78 &  58452.42
        \\
        NGC 4590 & 1 & 8400s & 92 & 76 & 58876.62\\
        \hline
    \end{tabular}
    \caption{The observational details of our GC survey, demonstrating multi-epoch observations for three of our GC targets.
    Stars were observed at multiple times across different epochs so as to mitigate the effect of binaries. During the third epoch, due to scheduling constraints, we split the targets of NGC~3201 into two exposures.}
    \label{tab:observation_log}
\end{table*}

\subsection{Data reduction}
\label{sec:data_reduction}

The raw data are primarily reduced with the \textsc{2dfdr}\footnote{\url{https://www.aao.gov.au/science/software/2dfdr}} pipeline \citep{2015ascl.soft05015A} default setting \textit{\tt aaomega1700D} provided by the AAT, which automatically subtracts the bias, calibrates the pixel-to-pixel sensitivity using the fibre flats, and calculates wavelengths with the arc lines. In addition, the \textsc{2dfdr} pipeline removes the sky spectrum, which is significant in the 1700D region.

The chemo-dynamical information, including radial velocity, are extracted using the CaT absorption lines. For this, we model each spectrum as consisting of the CaT lines and a continuum. The continuum is fit by means of a 6$^{\rm th}$-order polynomial with major spectral lines being masked out. We then normalise the flux of each spectrum to the best-fitting continuum. Then, we represent each line with a pseudo-Voigt profile (the summation of a Gaussian and a Lorentzian profile) as following: 
\begin{gather}
    \mathcal{F}(\lambda) = A_0 \mathcal{G}(\lambda,\lambda_0,\sigma_{\rm g}) + A_1 \mathcal{L}(\lambda,\lambda_0,\sigma_{\rm l}) \notag \\
    \mathcal{G}(\lambda,\lambda_0,\sigma_{\rm g}) = \frac{1}{\sqrt{2\pi}\sigma_{\rm g}}\mathrm{e}^{-(\lambda - \lambda_0)^2/(2\sigma_{\rm g}^2)} \notag \\
    \mathcal{L}(\lambda,\lambda_0,\sigma_{\rm l}) = \frac{\sigma_{\rm l}}{\pi((\lambda - \lambda_0)^2 + \sigma_{\rm l}^2)} \notag \\
    \lambda_0 = \lambda_{\textsc{lab}}\times(1 + z)
\end{gather}
where $z$ is the redshift, which is related to the velocity in the low-velocity regime through $z=v/c$.
The $A_{0}$ and $A_{1}$ parameters are the strength of the Gaussian and Lorentzian profiles respectively; $\lambda$ is the wavelength, and $\lambda_0$ is the spectral line centre; $\sigma_{\rm g}$ and $\sigma_{\rm l}$ indicate the line-width from the Gaussian and Lorentzian profiles respectively. The spectral template is constructed with three pseudo-Voigt profiles, whose line centres are correlated by the redshift.

We fit each spectrum with the CaT profile above. The data uncertainties come from the \textit{variance} from \textsc{2dfdr}, which are taken into account by convolving with the spectrum profile parameters' probability distribution. The best-fitting line profile parameters and their uncertainties (defined as the mean and the 1$\sigma$ quantiles) were derived by MCMC sampling of the posterior using {\sc emcee} \citep[][]{2013PASP..125..306F}. The systematic uncertainties are derived from the comparison between multiple epoch observations. The signal-to-noise (S/N) ratios are defined as the ratio of the absorption line strength to the residual surrounding the absorption lines ($\pm$5 \hbox{\AA}). Here we define a \textit{good\_star} when
\begin{gather}
    S/N > 3\ {\rm and} \notag \\
    \sigma_{\rm vlos} < 3\,\mathrm{km\,s^{-1}}.
\end{gather}
As for stars with multiple observations, those with a velocity difference larger than $5\,\mathrm{km\,s^{-1}}$ are clearly binaries and are excluded from the sample, otherwise we only adopt the velocity information from the spectrum that has the highest S/N. The equivalent width (EW) of each of the lines is calculated by integrating $\pm$ 20 \hbox{\AA} over the line centre and the uncertainties of the EW are derived by repeating the integration 100 times with random noise. Fig.~\ref{fig:sample} presents one example of a high S/N star and a zoomed-in view of the three lines as well as the best-fitting model. Tab.~\ref{tab:observation_log} lists the number of targets and good stars for each GC along with other observational information.

\begin{figure}
    \centering
    \includegraphics[width=\columnwidth]{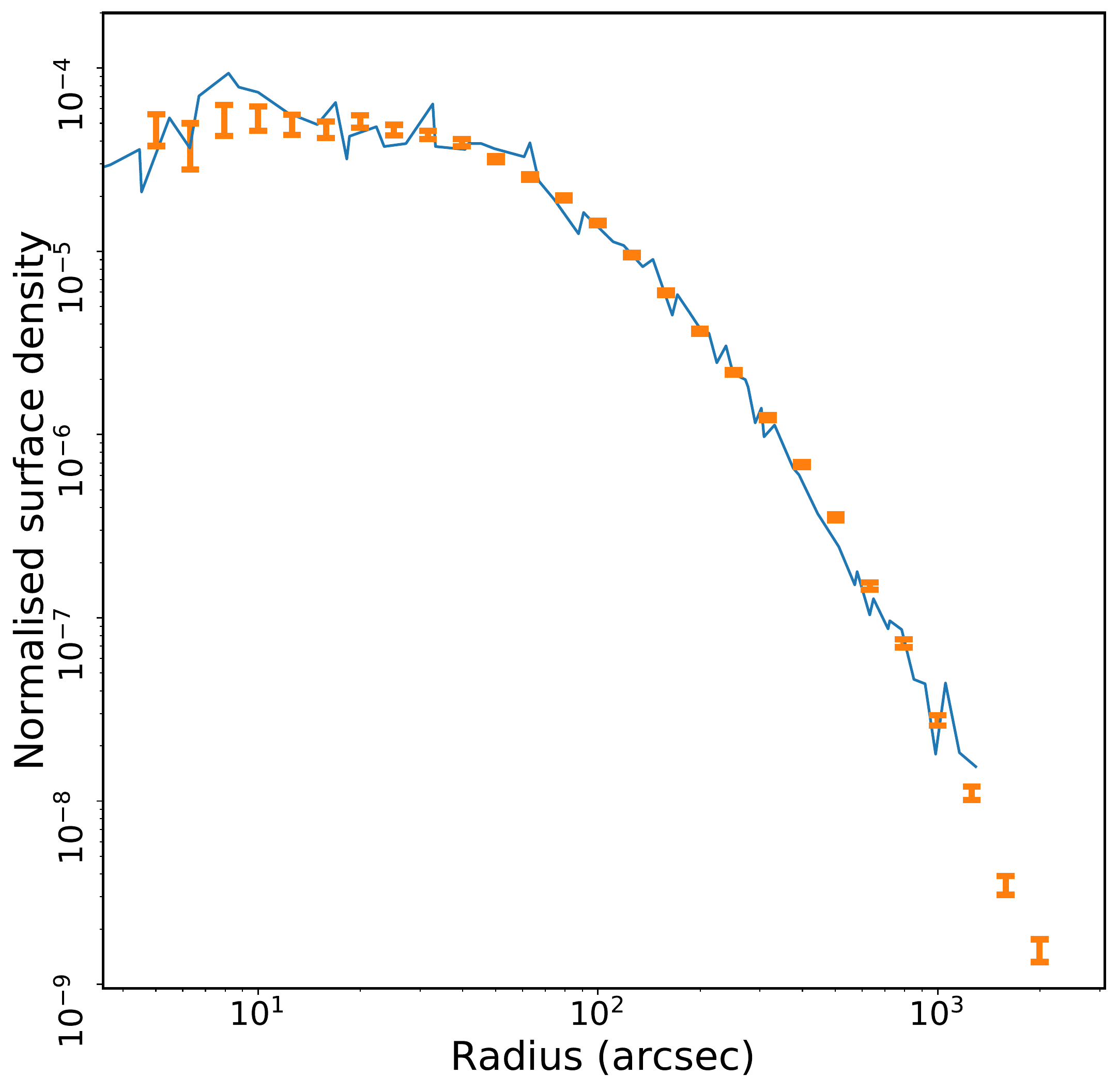}
    \caption{The number density profile of bright stars from the simulation (blue, solid line) compared to the observed surface brightness profile of NGC 3201 from \citet{1995AJ....109..218T} (orange errorbars). The simulation agrees excellently with the observed surface density profile.}
    \label{fig:surface_density}
\end{figure}

\begin{figure}
    \centering
    \includegraphics[width=\columnwidth]{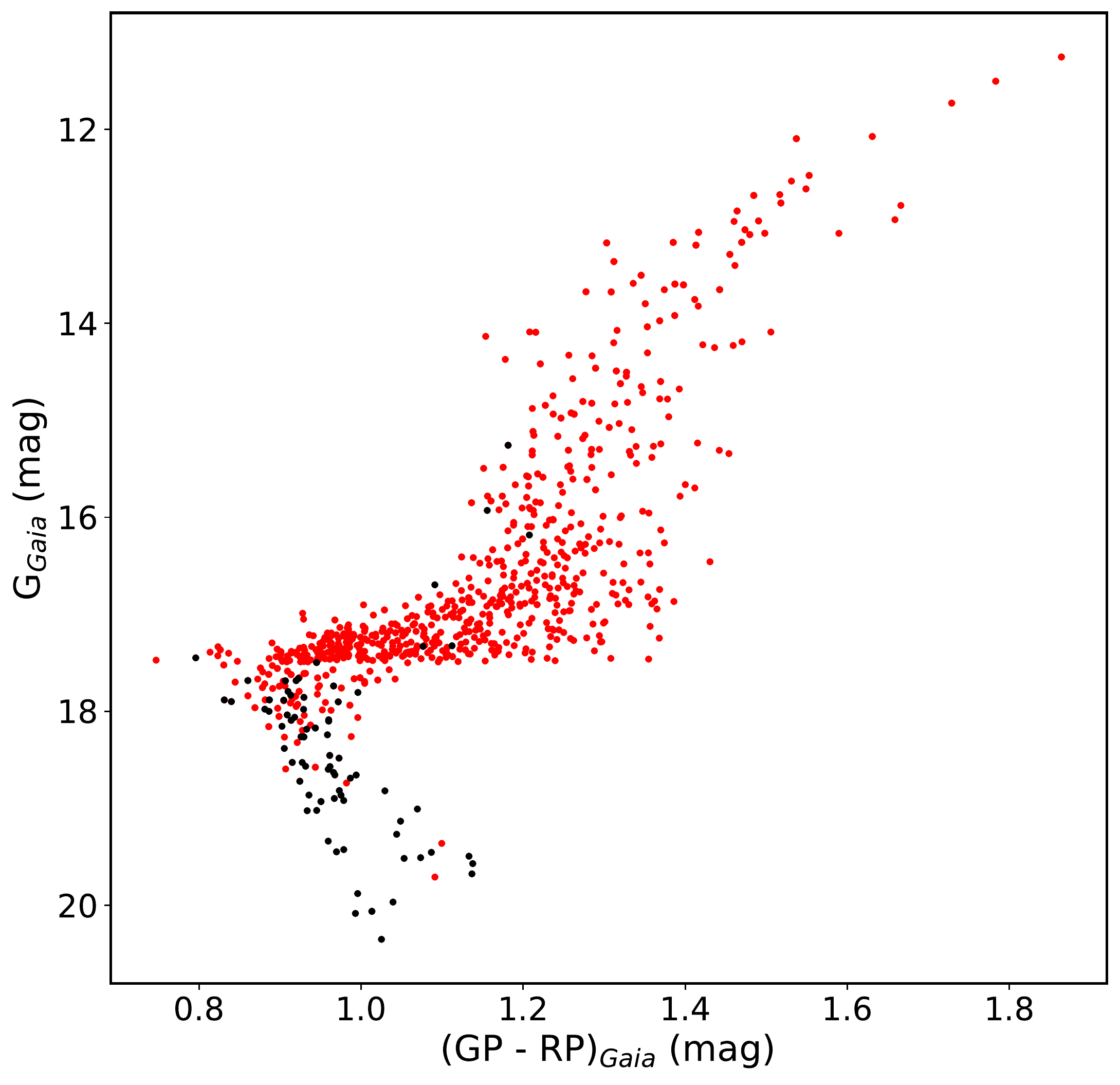}
    \caption{The \textit{Gaia} CMD of the NGC 3201 targets. The red dots are the \textit{good\_star} targets (see the definition in the text). The targets of the third epoch are brighter than $17.5\,\mathrm{mag}$, and most of the \textit{good\_star} targets are brighter than $18\,\mathrm{mag}$.}
    \label{fig:cmd}
\end{figure}

\begin{figure*}
    \centering
    \includegraphics[width=2\columnwidth]{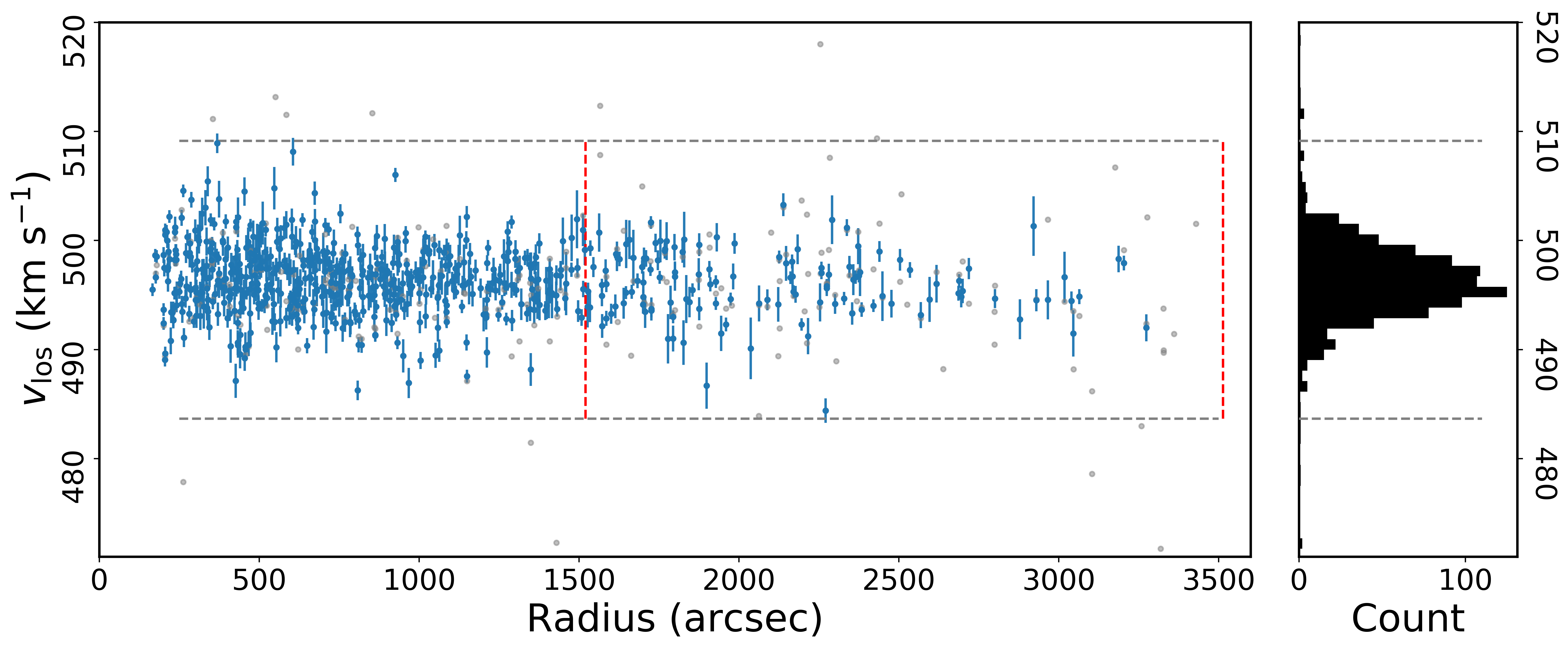}
    \caption{The LOS velocity of target stars, where the blue stars are those targets with S/N larger than 3, and grey dots are low S/N stars. Given the high radial velocity, the members of NGC~3201 are well separated from the contaminating MW halo stars. The right panel is the velocity distribution of all targets in the observation, which peaks at $496.4\,\mathrm{km\,s^{-1}}$. The two horizontal dashed lines indicate the $5\sigma$ range of the $v_{\mathrm{los}}$, and the two vertical red dashed lines indicate the King tidal radius \protect\citep{1996AJ....112.1487H} and the Jacobi radius \protect\citep{2018MNRAS.474.2479B} correspondingly.}
    \label{fig:velocity_radius}
\end{figure*}

\begin{figure*}
    \centering
    \includegraphics[width=2\columnwidth]{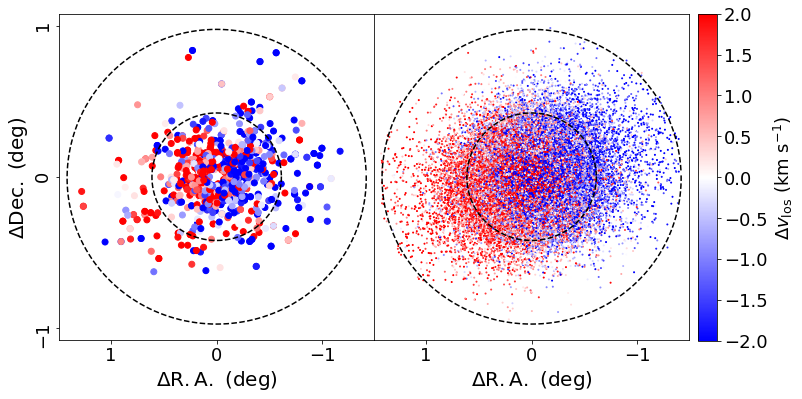}
    \caption{The projection on the sky of the observations (left panel) and the best-fitting simulation (right panel) colour-coded by their LOS velocities. The inner and outer dashed circles in both panels indicate the King tidal radius \protect\citep{1996AJ....112.1487H} and Jacobi radius \protect\citep{2018MNRAS.474.2479B}, respectively. The simulation exhibits a similar apparent rotational velocity pattern to the observed data, which is due to the effect of perspective view effect.}
    \label{fig:sky_vlos}
\end{figure*}

\begin{figure*}
    \centering
    \includegraphics[width=2\columnwidth]{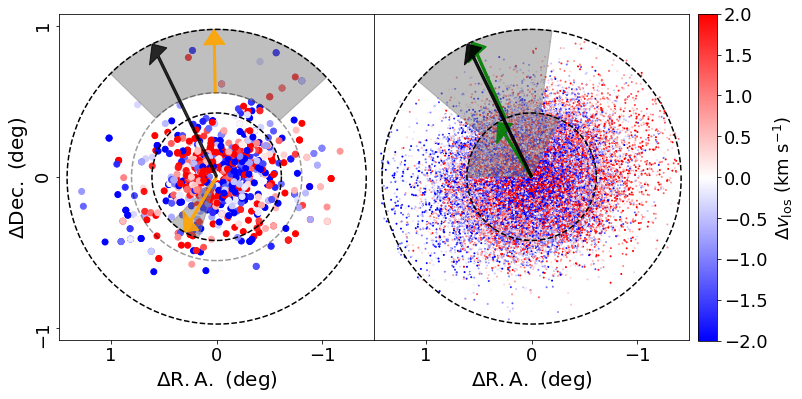}
    \caption{The comparison of LOS velocity with the perspective rotation effect subtracted between the observation (left panel) and simulation (right panel). The orange arrows (left) show the best-fitting PA for the stars with radius $433.2 < r < 956.7\,\mathrm{arcsec}$ and $r > 2000\ \mathrm{arcsec}$ (see the values in the Tab.~\ref{tab:dispersion_profile}) in the observations, and the grey region shows the $1\,\sigma$ range. Correspondingly, the green arrows (right) show the best-fitting PA for the stars in simulation, where the results from inner part (within King tidal radius) and outer part (beyond King tidal radius) are shown separately, and the grey regions again show the $1\,\sigma$ range. The black arrows in both panels indicate the expected rotational axis for the potential escapers (PA = $26.3^{\circ}$). Within the inner part, the disagreement between the simulation and observation suggests that this velocity variation comes from the internal rotation. At the outer part, there is some signal of unbounded stars that present the rotational direction aligned with the potential escapers.}
    \label{fig:internal_rotation}
\end{figure*}

\subsection{$N$-body simulations}
\label{sec:nbody}

In order to interpret the observation and estimate the influence of the external tidal field of the MW on the outer dynamical profile of NGC~3201, we performed a series of direct $N$-body simulations, which were made with the direct $N$-body code NBODY7 \citep{nitadoriaarseth2012} on the OzSTAR GPU cluster of Swinburne University and the GPU cluster of the University of Queensland. We have implemented the MW potential of \citet{Irrgangetal2013} as an additional option for an external tidal field in NBODY7, in order to model the influence of the MW on NGC~3201.

 
For our simulations, we first integrated the orbit of NGC~3201 backwards in time for 4~Gyrs in the MW potential of \citet{Irrgangetal2013} using a fourth-order Runge-Kutta integrator, with the initial phase space parameters from \citet{baumgardtetal2019}. We then set up a  $N$-body model of NGC~3201 that is non-rotating in an inertial reference frame and integrated the orbit of NGC~3201 forward in time to the present-day position using NBODY7. The initial $N$-body model was created based on the grid of $N$-body models described in \citet{2018MNRAS.478.1520B}, where the initial number of stars in the models of \citet{2018MNRAS.478.1520B} was 100,000 and do not contain primordial binaries. These models started from \citet{1962AJ.....67..471K} density profiles with varying concentration parameters $c$. The models of \citet{2018MNRAS.478.1520B} followed a range of initial mass functions, starting with those from \citet{2001MNRAS.322..231K} and extending towards those that are more strongly depleted in low-mass stars. The initial cluster models of \citet{2018MNRAS.478.1520B} were unsegregated, however mass segregation developed dynamically over time, so the simulations presented here started from already mass segregated models. This mass segregation increased further during the 4~Gyr duration of the simulations and developed into a cluster that is segregated in the same way as seen for NGC~3201. Since NGC~3201 loses about 5\% of its stars due to interaction with the tidal field of the MW during the 4 Gyrs of the simulation and also shrinks its core size by about 20\% due to the two-body relaxation driven evolution towards core collapse, we varied both the initial cluster mass and density profile slightly until we found the best match to the present-day observations of NGC~3201. Fig.~\ref{fig:surface_density} compares the surface density of bright stars in the simulation and the estimation based on the surface brightness from \citet{1995AJ....109..218T}. Also, comparing to a 2 Gyr simulation, we obtain similar results in terms of the final cluster size and the final velocity dispersion profile, hence we expect that an even longer simulation time will not change our final results. After the simulation finished, we extracted the particle data from the simulation, projected the cluster onto the sky and analysed the observations in the same way in which we analysed the observational data.

\section{Results and Discussion}
\label{sec:results}
As a result of the above observations and analysis for NGC~3201, we successfully extracted 886 good stellar spectra for 694 stars, and we excluded 11 clear binaries. Among the final sample, we have multiple epoch observations for 170 stars, and 94 stars (51 of them have multiple epoch observations) located beyond one half of the Jacobi radius, enabling us to characterise the dynamics out to a large distance from the cluster centre. Fig.~\ref{fig:cmd} presents the CMD of the cluster, where the \textit{good\_stars} are colour-coded in red. At $V=18\,\mathrm{mag}$ the number of good stars decreases significantly. The mean LOS velocity of NGC~3201 is $496.47\pm0.11\,\mathrm{km\,s^{-1}}$ based on the observations. A summary of these results is presented in Fig. \ref{fig:velocity_radius}, which shows the LOS velocity distribution for the target stars, including their individual values as a function of cluster-centric radius. With this, we address several key scientific questions about NGC~3201.

\subsection{Is NGC~3201 rotating?}
\label{sec:rotating}

One of the most prominent dynamical signatures would be rotation, which is a record of the cumulative effects from the birth of the GC, two-body relaxation and interaction with the tidal field of the host galaxy. A good understanding of the rotation is also important for determining the velocity dispersion \citep[e.g.][]{1995ApJ...454..788C,2018ApJ...860...50F,2019MNRAS.485.1460S,2012A&A...538A..18B}.  With precise LOS velocity measurements, internal rotation has been found in different clusters \citep[e.g.][ and references therein]{2018MNRAS.473.5591K,2018MNRAS.481.2125B, 2019MNRAS.485.1460S} and is typically measured using the LOS velocity difference on both sides of a central axis. The left panel of Fig.~\ref{fig:sky_vlos} shows the tangent plane projection of stars in  NGC~3201 coloured by the measured LOS velocities (with the mean LOS velocity subtracted), where the inner and outer dashed circles indicate the King tidal radius \citep[36.1 pc, or 1520 arcsec,][]{1996AJ....112.1487H} and an estimation of the Jacobi radius  \citep[83.46 pc, or 3513 arcsec,][]{2018MNRAS.474.2479B}, respectively. Similarly, the right panel shows the best-fitting simulation, colour-coded with the LOS velocity. 

The velocity variation of member stars in NGC~3201 can be reproduced by the best-fitting simulation. As shown in Fig.~\ref{fig:sky_vlos}, there is an obvious rotation signal with amplitude of $\sim5\,\mathrm{km\,s^{-1}}$, with stars in the east (positive $\Delta$R.A.) are moving away from the observer (relative to the cluster centre), while stars in the west (negative $\Delta$R.A.) are moving towards us (relative to the cluster centre). 
This signal is mostly due to perspective rotation, which is important for objects with a large angular diameter that have a high systemic proper motion. For the distance, systemic proper motion, and diameter of NGC~3201 ($\sim5\,$kpc, $\sim10\,\mathrm{mas\,yr^{-1}}$, $\sim100\,$arcmin respectively) the magnitude of this effect is a velocity difference of $\sim 7\,\mathrm{km\,s^{-1}}$ \citep{2006A&A...445..513V}. Perspective rotation is therefore important and responsible for most of the signal seen in Fig.~\ref{fig:sky_vlos}.

In addition to the perspective rotation effects, several other scenarios can also lead to rotation. For example, as the GC evolves within the MW potential the LOS velocity varies along the GC orbit, which will be most obvious for the unbound stars in the tidal tails. Furthermore, potential escapers at large radii can contribute to the rotation signal. Finally, internal rotation from the formation of the GC could naturally lead to an observable rotation.

To calculate the perspective rotation effect on the observed LOS velocities of NGC~3201, we assume that NGC~3201 is centred on $(R.A., Dec.) = (154.40^{\circ}, -46.41^{\circ})$ at a heliocentric distance of $4.9$ kpc \citep{Harris10}, with a systemic LOS velocity of $496.47\,\mathrm{km\,s^{-1}}$ and proper motion of $(\mu_{\alpha}^{\*}, \mu_{\delta}) = (8.37, -1.96)\,\mathrm{mas\,yr^{-1}}$. We calculate the systemic velocity of the simulation by taking the mean of the stars within 0.5 degree around the GC centre. The perspective rotation effect can then be calculated for each star from the systemic velocity using Eq.~6 of \citet{2006A&A...445..513V}. To adjust for this effect in the observed and simulated stars, it is then subtracted from each star's corresponding observed LOS velocity. Fig.~\ref{fig:internal_rotation} shows the LOS velocity of the simulation and the observation after having adjusted for the perspective rotation effect.

A simple relation that includes a rotation component:
\begin{gather}
    v_{\mathrm{los}, 0} = A_{\mathrm{rot}} \sin (\phi - \phi_0) + v_{\mathrm{sys}} \notag \\
    p(v_{\mathrm{los}}) = \frac{1}{\sqrt{2\pi}\sigma_{v_{\mathrm{los}}}} \exp^{-\frac{(v_{\mathrm{los}} - v_{\mathrm{los}, 0})^2}{2\sigma_{v_{\mathrm{los}}}^2}}
    \label{eq:model1}
\end{gather}
is fitted to the residual velocity within radial bins for both observation and simulation including the velocity errors in quadrature. Here $A_{\mathrm{rot}}$ is the amplitude of the velocity difference; $\phi$ is the directional angle from the rotation axis increasing from the north to the east and $\phi_0$ is the reference positional angle (PA); $v_{\mathrm{sys}}$ is the systemic velocity along the line-of-sight and $\sigma_{v_{\mathrm{los}}}$ is the intrinsic LOS velocity dispersion. The posterior parameters space is sampled with an MCMC approach, and the best-fitting parameters and $1\sigma$ uncertainties are summarised in Tab.~\ref{tab:dispersion_profile}. In the inner part of the cluster (for stars within $\sim900$ arcsec around the GC centre), we found a signal of rotation with amplitude of around $1\,\mathrm{km\,s^{-1}}$ (see Tab.~\ref{tab:dispersion_profile} for detailed profile). The amplitude of the rotation becomes weaker at larger radius. For stars beyond $900$ arcsec radius, the amplitude decreases to $A_{\mathrm{rot}} \approx 0.35^{+0.32}_{-0.24}\,\mathrm{km\,s^{-1}}$. The PA of the rotation axis in the inner part is $\sim 133^{\circ}$ -- $170^{\circ}$, which is significantly different from the simulation (PA $= {30.3^{\circ}}_{-68.1^{\circ}}^{+62.6^{\circ}}$ with a very weak amplitude $ A_{\rm rot} = 0.13^{+0.11}_{-0.09}\,\mathrm{km\,s^{-1}}$ for stars with radius between 300 arcsec and 900 arcsec, and PA $= {22.5^{\circ}}_{-30.7^{\circ}}^{+27.6^{\circ}}$ with an amplitude $A_{\mathrm{rot}} = 0.64^{+0.33}_{-0.32}\,\mathrm{km\,s^{-1}}$ for stars between the King tidal radius and the Jacobi radius). At the outer part of the GC, where the radius $r > 2000\,\mathrm{arcsec}$, we found an opposite rotational signal compared to the rotation at the inner part, with $A_{\mathrm{rot}} = 0.80^{+0.49}_{-0.47}\,\mathrm{km\,s^{-1}}$ and PA $= {0.6^{\circ}}_{-49.2^{\circ}}^{+45.8^{\circ}}$, which agrees with the rotational direction of the simulation. This counter-rotation at the outermost region relative to the inner one is in good agreement with the prediction of a tidally perturbed, rotating stellar cluster from \citet{2018MNRAS.475L..86T}. The visualisation of the results is presented in the Fig.~\ref{fig:internal_rotation}.

\begin{figure}
    \centering
    \includegraphics[width=\columnwidth]{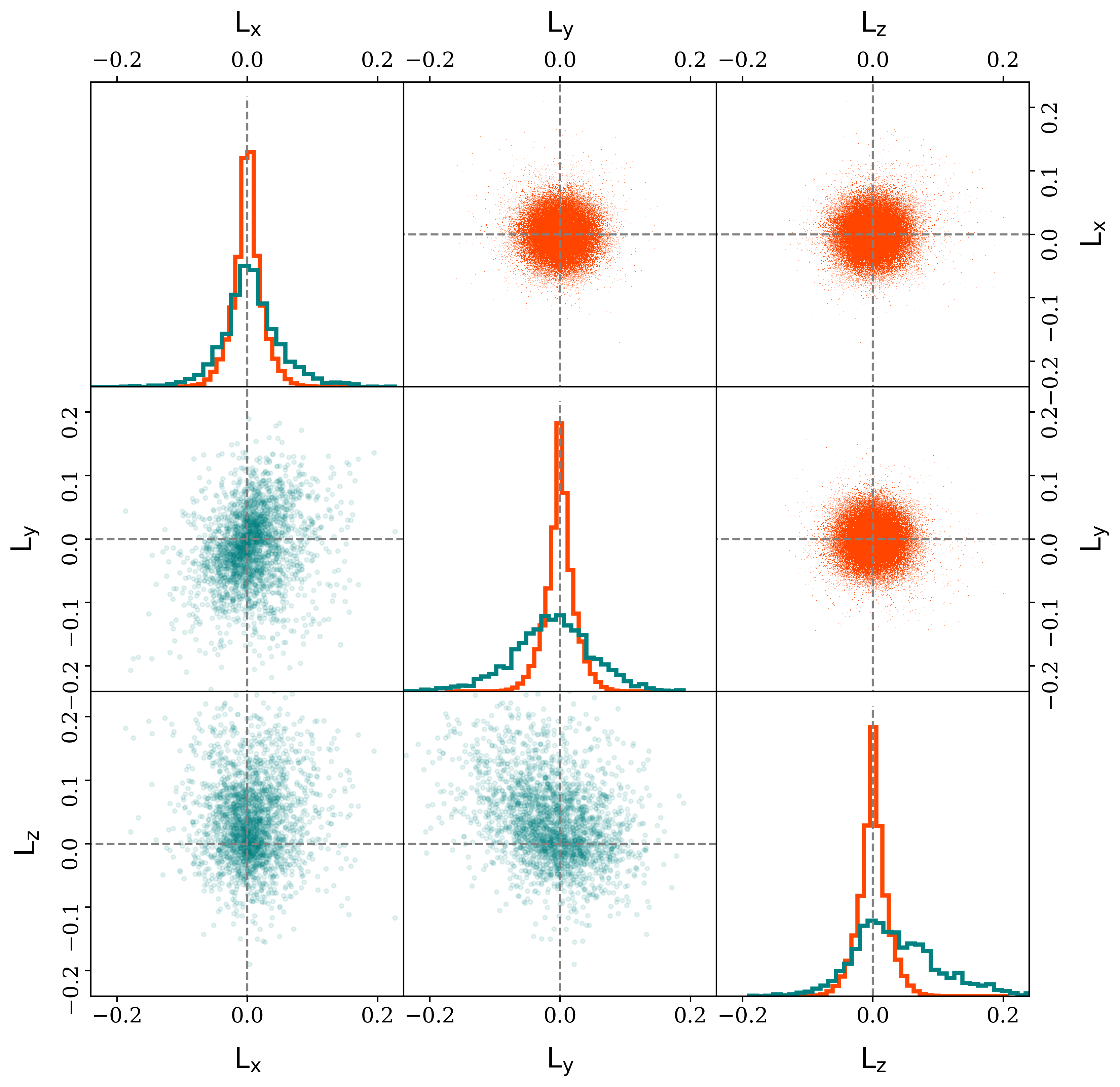}
    \caption{The components of the angular momentum of stars of the simulation with $r < 2000\,\mathrm{arcsec}$ (orange-red) and $2000\,\mathrm{arcsec} < r < 3600\,\mathrm{arcsec}$ (green) from the GC centre. The inner part shows no significant angular momentum signal, suggesting that the GC does not have internal rotation. At the outer part, the angular momentum bias comes from the potential escapers.}
    \label{fig:angular_momentum_corner}
\end{figure}

The rotation could be characterised by angular momentum relative to the GC centre. Though we do not have the full phase-space information for the observation data, we can explore the dynamical features of the simulation. Hence, we calculated the present-day angular momentum of the stars in the simulation with respect to the GC centre. Here we use the coordinate system that has the $z$ axis perpendicular the orbital plane, and the $x$ axis aligned with the systemic velocity of the GC, but the frame is inertial, i.e. non-rotating. Fig.~\ref{fig:angular_momentum_corner} presents the angular momentum of stars---with $r < 2000\,\mathrm{arcsec}$ and $2000\,\mathrm{arcsec} < r < 3600\,\mathrm{arcsec}$ around the GC centre. For all stars at the inner part, the mean and dispersion of each components  of the specific angular momentum are $(L_x,L_y,L_z) = (0.00 \pm 0.02, 0.00 \pm 0.02, 0.00 \pm 0.02)\,\mathrm{kpc\,km\,s^{-1}}$; for stars at the outer part, we find $(L_x,L_y,L_z) = (0.01 \pm 0.06, -0.01 \pm 0.06, 0.04 \pm 0.06)\,\mathrm{kpc\,km\,s^{-1}}$. The results at the inner part indicates that the GC has no internal rotation after 4 Gyr evolution, and the clear bias from zero at the outer part is due to the potential escapers. We can compare these values to what is expected from potential escapers. For circular orbits, in a reference frame that co-rotates with the orbit, prograde stars are preferentially lost, resulting in a net retrograde solid-body rotation of potential escapers \citep{2017MNRAS.466.3937C, 2017MNRAS.468.1453D}. \cite{2016MNRAS.461..402T} showed the average angular frequency of the potential escapers is $\langle\Omega_{\rm PE}\rangle  =-0.5\Omega_{\rm orb}$, where $\Omega_{\rm orb}$ is the angular frequency of the orbit. In a non-rotating frame $\langle\Omega_{\rm PE}\rangle  = +0.5\Omega_{\rm orb}$. We can approximate the eccentric orbit of NGC~3201 by a circular orbit at Galactocentric radius $R_{\rm p}(1+e)\simeq13~$kpc \citep{2003MNRAS.340..227B,2016MNRAS.455..596C}, where $R_{\rm p}\simeq9~$kpc is the pericentre distance and $e\simeq0.5$ the eccentricity \citep{2018A&A...616A..12G}. Assuming a flat rotation curve of 220~km/s, $\Omega_{\rm orb}\simeq 0.017/$Myr and thus $\Omega_{\rm PE}\simeq8.7\times10^{-3}/$Myr. The average orbital velocity at $0.75r_{\rm Jacobi}\simeq 63~$pc is then $0.55~$km/s, and in the non-rotating frame the average angular momentum is $\langle L_z\rangle\simeq0.035~\mathrm{kpc\,km\,s^{-1}}$, i.e. as we find in the $N$-body model, suggesting that the rotation we see in the outskirts of the $N$-body model is due to potential escapers.

The rotation axis of the potential escapers is aligned with the angular momentum vector of the Galactic orbit, which is well-constrained by the systemic proper motion, distance and line-of-sight velocity of NGC~3201. In Fig.~\ref{fig:internal_rotation} we show in both panels with a black arrow the projection of the angular momentum vector of the Galactic orbit. 
As Fig.~\ref{fig:internal_rotation} shows, the direction of the rotational axes in the inner part of the observation indicates that the rotational signal within $r < 900\,\mathrm{arcsec}$ (the orange arrow) is different from what is expected from potential escapers (the black arrow). Hence this signal at the inner part of NGC~3201 is likely to be the internal rotation of the GC. At the outer part ($r > 2000\,\mathrm{arcsec}$), we found that the direction of the rotational signal is aligned with the potential escapers, which suggests that those stars at the outskirt are likely energetically unbound, yet associated to the cluster \citep{1970A&A.....9...24H}. As far as we are aware, this is the first detection of this predicted signal of potential escapers in a star cluster. In the $N$-body simulation, the rotation in both inner and outer regions aligns with the potential escaper prediction. This is because the model started without rotation, so all the rotational signal is imposes by the tides.

\begin{figure}
    \centering
    \includegraphics[width=\columnwidth]{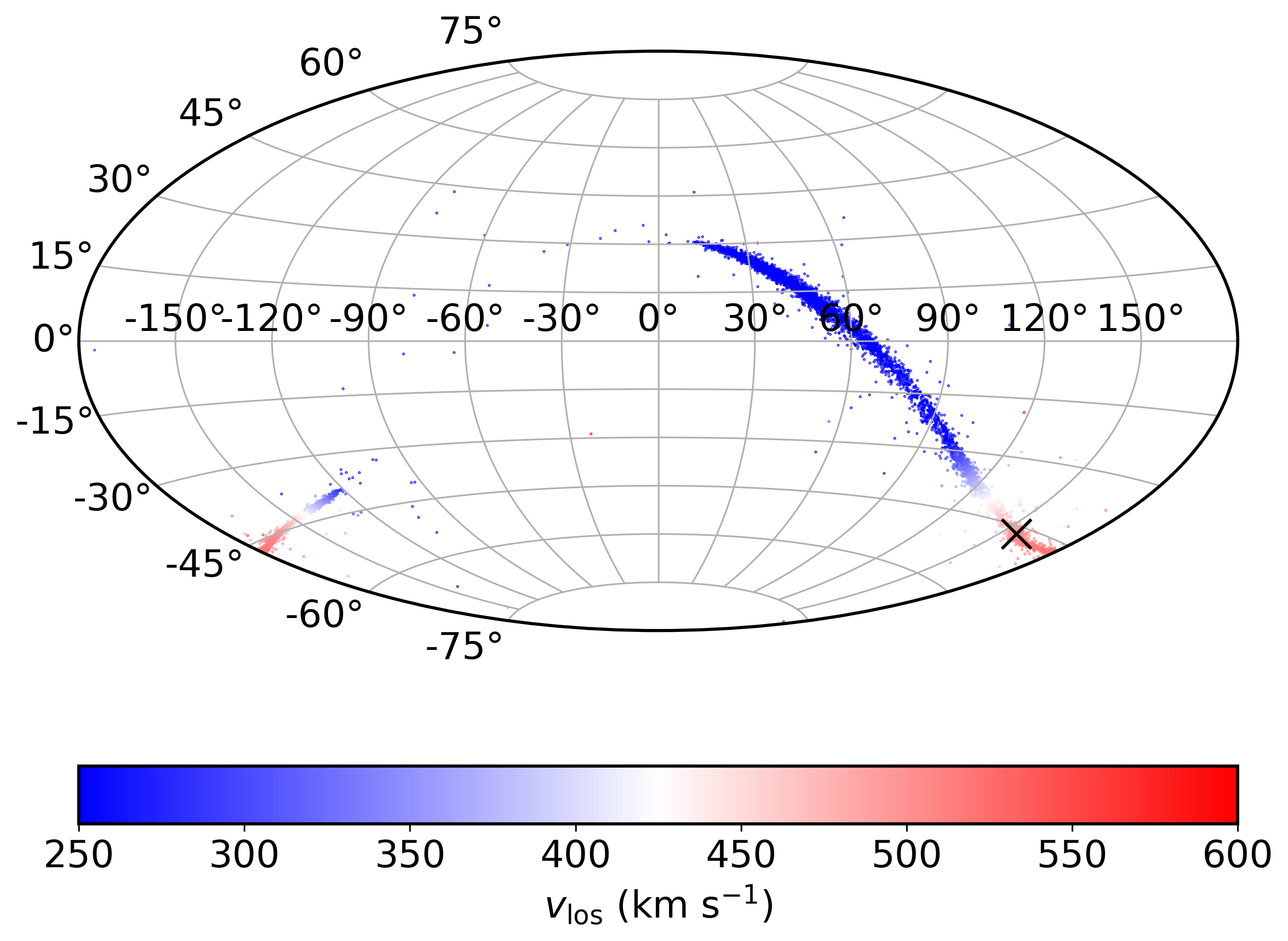}
    \caption{An all-sky Aitoff projection of the output of the simulation of NGC~3201, colour-coded with respect to the LOS velocity. The cross indicates the location of the GC. The plot shows that the change of the LOS velocity is not a local effect, but extends continually along the tidal arms around the MW. The LOS velocity of the GC is close to the maximum, and decreases significantly along the stream. At distances larger than 5 degree from the centre of NGC~3201, the LOS velocity is smaller than $480\,\mathrm{km\,s^{-1}}$.}
    \label{fig:GC_orbit}
\end{figure}

\subsection{Velocity dispersion}
For a pressure-supported system in dynamical equilibrium, the dispersion is directly relates to the average internal kinetic energy. This is then related to gravitational potential energy as based upon the virial theorem. Hence the mass profile of the stars, as well as any dark content, can in principle be estimated by measuring the dispersion profile. A typical way of interpreting the result is to compare the measured profile to a model \citep[e.g.][]{2019ApJ...887L..12B, 2019MNRAS.484.2832V,2019MNRAS.483.1400H}. In this section, we present our estimate of the dispersion profile with a higher precision and discuss the effects of the MW potential, binaries and the escape rate (extra-tidal stars, considering the effect of stellar-mass BHs).

\begin{figure*}
    \centering
    \includegraphics[width=2\columnwidth]{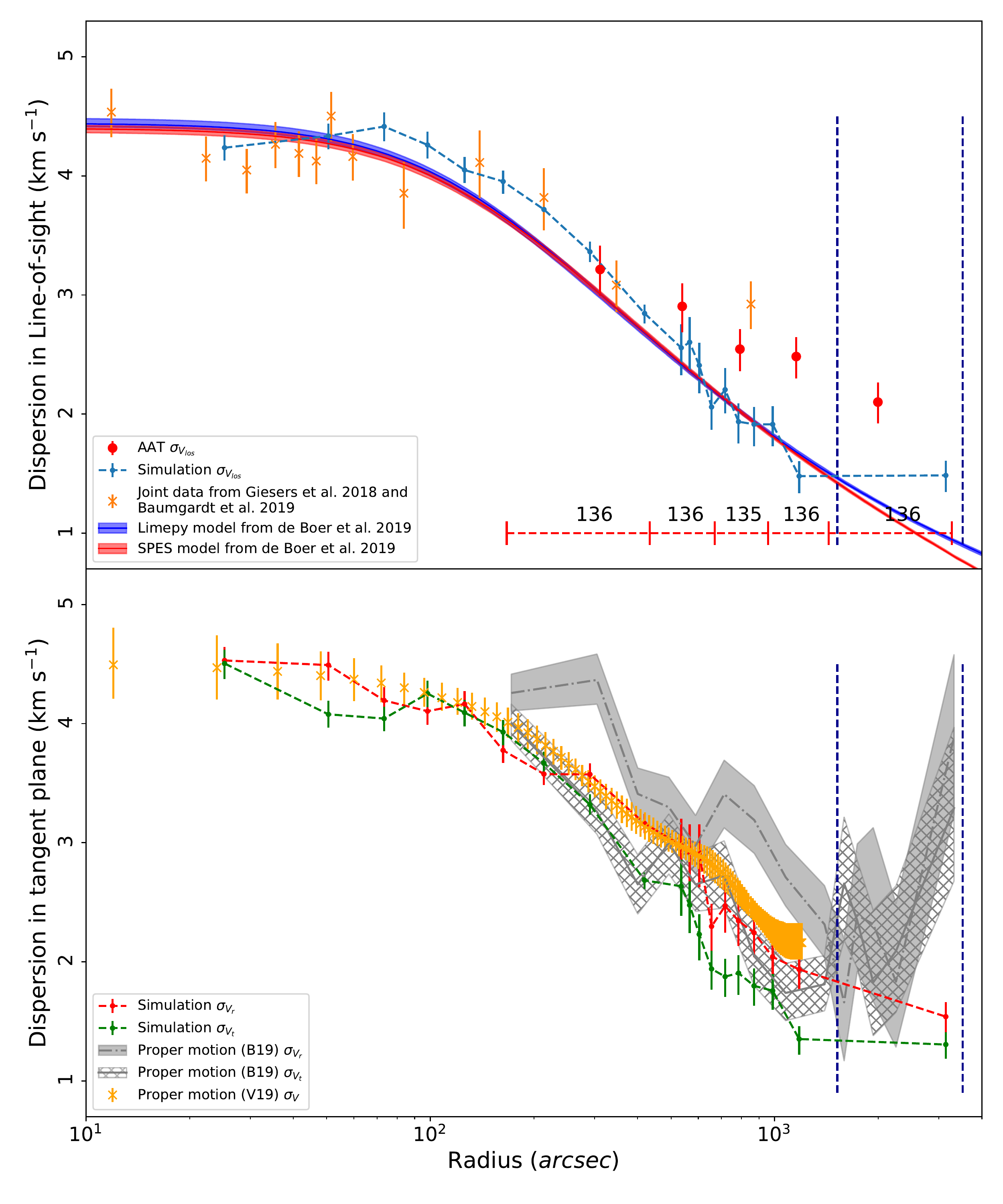}
    \caption{The top panel depicts the dispersion in the LOS from the N-body model (blue), from the AAT 2dF/AAOmega observations (red), and from the previously published data (orange). The {\sc limepy} and {\sc spes} models with $1\sigma$ uncertainties from \citet{2019MNRAS.485.4906D}  (blue and red regions) are also presented, where the mass of NGC~3201 is 27\% smaller than the mass from \citet{baumgardtetal2019}, which leads to a best fit to the previously published data \citep{2018MNRAS.478.1520B,2019A&A...632A...3G}. The red dash lines at the bottom indicate the radial range of each bin and the number above them are the number of stars in each bin. The errorbars indicate the $1\sigma$ uncertainty. The bottom panel depicts the dispersion in the tangent plane from the N-body model, and from \protect\citet[][V19]{2019MNRAS.489..623V} and \protect\citet[][B19]{2019ApJ...887L..12B} (including the radial (grey) and tangential (grey hatched) components, where the regions indicate the $1\sigma$ ranges respectively). In both panels, the left and right dark vertical dashed lines mark the King tidal radius and Jacobi radius respectively. Within $r = 500\ \mathrm{arcsec}$, the simulation agrees well with observation, however the observed dispersion is significantly larger than that in the simulation at large radii.}
    \label{fig:dispersion_profile}
\end{figure*}

As noted in the previous section, NGC~3201 presents a pattern where the LOS velocities are larger on the east side than on the west side. This systematic variation of velocity has to be taken into consideration when estimating the dispersion. Here, the dispersion was included in the rotation model (equation~3), and the best-fitting intrinsic dispersions are listed in  Tab.~\ref{tab:dispersion_profile}. As a consistency check and to demonstrate the dispersion profile in the inner part of the GC, we also included previously published $v_{\mathrm{los}}$ data \citep{2018MNRAS.478.1520B,2019A&A...632A...3G}, and the proper motion dispersion profiles from \citet{2019ApJ...887L..12B} and \citet{2019MNRAS.484.2832V}. Meanwhile, we fit the same relation to the $N$-body simulation described in Sec.~\ref{sec:nbody}, and include the dispersion profiles from the {\sc limepy} models \citep{2015MNRAS.454..576G} and {\sc spes} models \citep{2019MNRAS.487..147C} from \citet{2019MNRAS.485.4906D} as comparisons. 

Fig.~\ref{fig:dispersion_profile} shows the velocity dispersion profile from the observations and simulation, where the top panel shows the dispersion in LOS, and the bottom panel shows the dispersion in the tangent plane. The two dashed lines in both panels indicate the King tidal radius and Jacobi radius, respectively. The velocity dispersion within $\sim 500\,\mathrm{arcsec}$ from the GC centre can be well reproduced by the simulation, and our observations agree with the data from the literature. However, the simulation is significantly lower than the observations at larger radius. At radii beyond $2000\,\mathrm{arcsec}$, we find that the LOS velocity dispersion is $\sim 2.01\pm0.18\,\mathrm{km\,s^{-1}}$, and tends to flatten outwards, whereas the dispersion of the simulation decrease faster with radius and is about $1.48\pm0.14\,\mathrm{km\,s^{-1}}$, i.e. $\sim2\,\sigma$ lower. The dispersion profiles of the {\sc limepy} and {\sc spes} models are presented as red and blue regions respectively. Compared to \citet{1966AJ.....71...64K} models, the{\sc limepy} models have an additional degree of freedom that describes the `sharpness' of the energy truncation, and these models are therefore more flexible in describing the outer density profiles. The {\sc spes} model include a prescription for potential escapers.

The model parameters are taken from \citet{2019MNRAS.485.4906D}. The cluster mass is 27\% smaller than the mass from \citet{baumgardtetal2019}, so that the models have the best fit to the inner part data from \citet{2018MNRAS.478.1520B,2019A&A...632A...3G}. We note that the $N$-body models from \citet{baumgardtetal2019} are multi-mass, where the massive stars move a bit slower due to equipartition, which allows for a larger mass than the {\sc limepy} and {\sc spes} models. The dispersion profiles from these two models agree with the $N$-body simulation. However, both models underestimate the dispersion of the GC at the outer part. As for the dispersion in the tangent plane, the dispersion profile from \citet{2019MNRAS.484.2832V} agrees with the tangential dispersion profile from \citet{2019ApJ...887L..12B}. The tangential dispersion is lower than the radial dispersion profile, which is expected if the cluster has radially biased velocity anisotropy. Our results agree well with the observations in the inner part of the GC, especially, the dispersion profile from \citet{2019MNRAS.484.2832V} and the tangential dispersion profile from \citet{2019ApJ...887L..12B}. In the outer part of the GC, the observed proper motion dispersion in the tangent plane is significantly larger than the simulation.  Similarly, \citet{2019ApJ...887L..12B} compare the proper motion dispersion profiles to the model for the dispersion of potential escapers from \citet{2017MNRAS.466.3937C}, finding that the observed dispersion out to the Jacobi radius is approximately half the model prediction of \citet{2017MNRAS.466.3937C}. This is perhaps not too surprising, because the potential escaper model of \citet{2017MNRAS.466.3937C} was derived for circular orbits, and NGC\,3201 is near pericentre, where the dispersion of potential escapers is about twice as high as near apocentre for an eccentricity of 0.5 \citep[see figure 10 in][]{2019MNRAS.487..147C}. 
However, the potential escapers are present in the $N$-body simulation and their effects are included in the dispersion profile of the simulation. Hence the potential escapers are unlikely to cause the observed large dispersion.

Several scenarios could potentially explain the discrepancy between observations and simulations. The large dispersion might relate to the interaction with the galactic potential. The heating when the GC crosses the galactic disk might also increase the dispersion \citep[e.g. $\omega$ Cen, ][]{2012ApJ...751....6D}. However, compared to $\omega$ Cen, NGC~3201 has a much larger peri-galacticon radius \citep{baumgardtetal2019}, where the disk heating is insignificant. In addition, the $N$-body model we adopted includes the influence from the MW (as well as the disk). We conclude that the excess dispersion is unlikely to be a result of the interaction between the GC and the potential field of the MW.

In the following sections, we will discuss two additional mechanisms that might lead to the flattened dispersion profile in the outer parts of NGC~3201, and our estimation of their effects on the observations.

\begin{table*}
    \centering
    \begin{tabular}{c c c c c c c}
        \hline 
        $R_{\mathrm{low}}$ & $R_{\mathrm{high}}$ & $<R>$ & $A_{\mathrm{rot}}$ & $\phi_0$ & $\sigma_{\mathrm{los}}$ &  N \\
        $(\mathrm{arcsec})$ & $(\mathrm{arcsec})$ & $(\mathrm{arcsec})$ & $(\mathrm{km\,s^{-1}})$ & $^{\circ}$ & $(\mathrm{km\,s^{-1}})$ & \\
        \hline 
        $166.6$ & $433.2$ & 312.5 & $0.39^{+0.37}_{-0.27}$ & $89.9_{-73.4}^{+64.8}$ & $3.21^{+0.22}_{-0.20}$ & 136 \\
        $433.2$ & $669.3$ & 544.8 & $0.88^{+0.41}_{-0.42}$ & $169.8_{-24.4}^{+26.9}$ & $2.90^{+0.22}_{-0.19}$ & 136 \\
        $669.3$ & $956.7$ & 804.0 & $1.09^{+0.37}_{-0.37}$ & $132.4_{-16.6}^{+18.5}$ & $2.54^{+0.19}_{-0.17}$ & 135 \\
        $956.7$ & $1434.6$ & 1181.0 & $0.35^{+0.32}_{-0.24}$ & $49.3_{-80.8}^{+50.6}$ & $2.48^{+0.19}_{-0.16}$ & 136 \\
        $1435.6$ & $3273.8$ & 2019.0 & $0.51^{+0.35}_{-0.32}$ & $312.1_{-33.4}^{+35.6}$ & $2.01^{+0.18}_{-0.16}$ & 136 \\
        \hline 
        $433.2$ & $956.7$ & 665.3 & $0.96^{+0.26}_{-0.26}$ & $149.8_{-14.3}^{+15.5}$ & $2.70^{+0.14}_{-0.13}$ & 272\\
        $2000$ & $3600$ & 2468 & $0.80^{+0.49}_{-0.47}$ & $0.6_{-49.3}^{+45.8}$ &  $2.40^{+0.33}_{-0.29}$ & 54\\
        \hline
        
    \end{tabular}
    \caption{The estimated velocity dispersion profile. The first and second columns shows the range in radius; the third column give the mean radius in each bin; the fourth column shows the estimated dispersion and $1\sigma$ uncertainties; and the last column gives the number of stars within each bin. The last two rows present the fitting results of the rotational signal from the inner and outer parts of the GC respectively.}
    \label{tab:dispersion_profile}
\end{table*}

\begin{figure}
    \centering
    \includegraphics[width=\columnwidth]{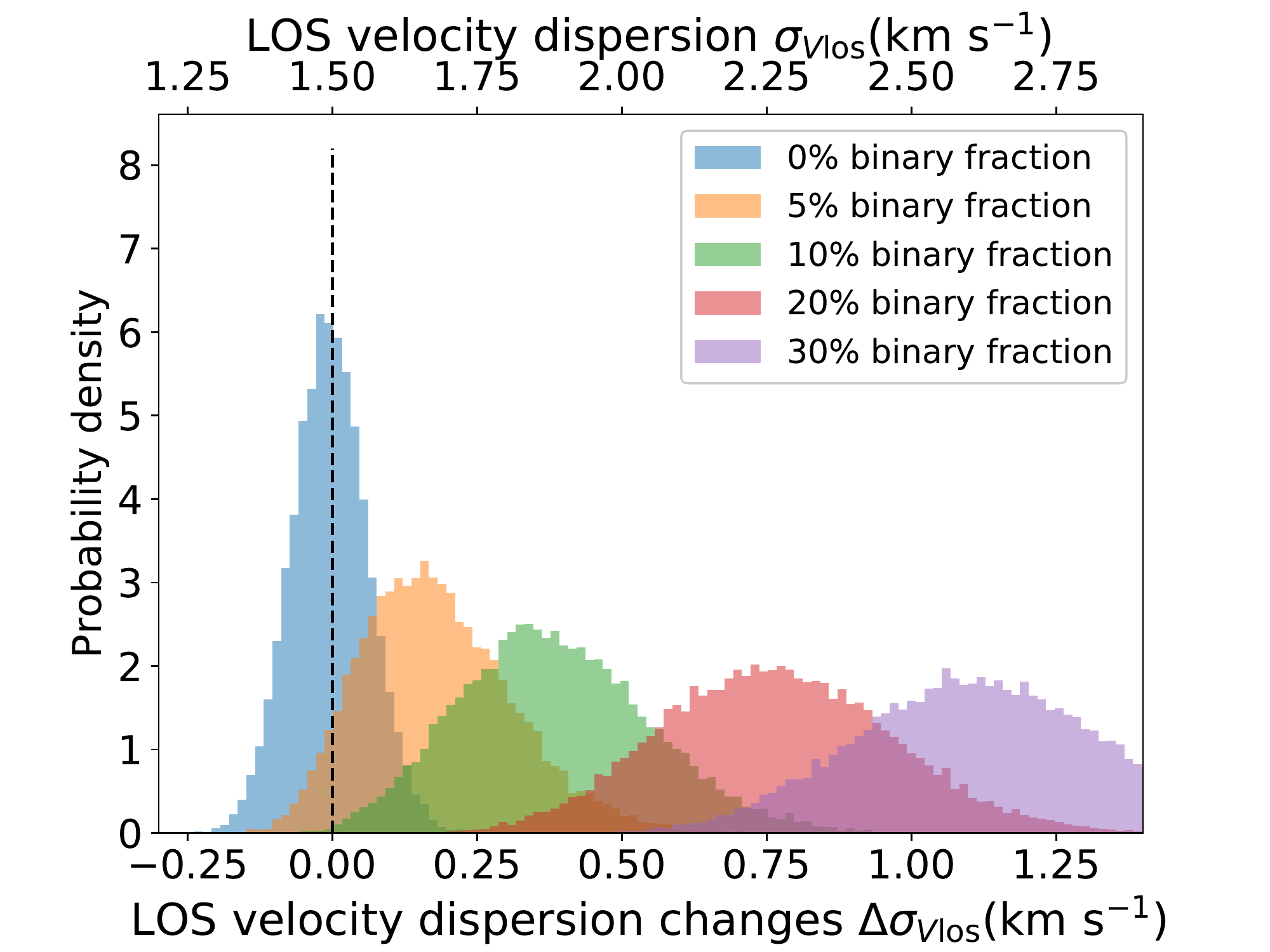}
    \caption{The effect of binaries on the velocity dispersion profile. With the initial dispersion set to be $1.5\,\mathrm{km\,s^{-1}}$, this figure shows the distribution of the final dispersion. Given a reasonable binary fraction of less than 15\%, the presence of binaries cannot significantly change the estimated dispersion.}
    \label{fig:binary_dispersion}
\end{figure}

\subsection{The effect of binaries}
The observed velocity of a binary can be significantly different from the systemic velocity of the GC due to its internal orbital velocity. Some binaries with a large velocity deviation from the GC velocity or with a short period would be easily identified (see Fig.~\ref{fig:N3201_footprint} and \ref{fig:velocity_radius}), while some long period binaries are difficult to detect. The presence of undetected binaries would inevitably influence the measured velocity and hence the estimated dispersion. 

The velocity of a star in a binary system varies periodically, and with our multi-epoch observation, short-period binaries with a significant velocity difference at different epochs are easily identified as binaries. Fig.~\ref{fig:N3201_footprint} shows the velocity difference between epochs, where some binaries deviate significantly from zero. Long-period binaries, however, are difficult to identify. We also note that a fraction of our targets were only observed for a single epoch, and binaries (even short-period) within that sub-sample cannot be identified. Hence, a proper estimation of the effect of binaries is necessary for the dynamical analysis.

To estimate the effect of binaries, we first randomly sampled binaries with \textsc{velbin} \citep{2012A&A...547A..35C,2014A&A...562A..20C} from distributions of period, mass ratio and eccentricities appropriate for solar-type binaries \citep{2010ApJS..190....1R}. Given that our target stars are either near the main-sequence turnoff or RGB stars, we assume a mass of $0.8\,\mathrm{M_{\odot}}$ for the primary star in the binary systems. Since softer binaries would be disrupted in a cluster environment, we retained only hard binaries. \citet{2019A&A...632A...3G} found that all the binaries for which they secured orbital solutions in NGC 3201 have energies a factor of $\sim5$ or more above the hard-soft boundary.   We therefore kept only hard binaries with an orbital velocity larger than three times the current central velocity dispersion of the cluster. This translates into a higher minimum period for the binaries, and accounts for the possibility that binaries with an energy just above the present-day hard-soft boundary have been destroyed in the past when the cluster was more massive and more compact.

Based on the binary sample, we constructed mock radial velocity data sets with time baselines, radial velocity uncertainties, and numbers of epochs that mimic our observed cluster member stars in the two outer bins shown in Fig.~\ref{fig:dispersion_profile}. The cluster velocity dispersion is initialised to be $1.5\,{\mathrm{km\,s^{-1}}}$, comparable to the velocity dispersion of the best-fitting simulation in the outermost radial bins. We adopted different binary fractions and just like in the real observations we excluded from the final data sets the stars that would have been identified as binaries, as well as stars that would have not been retained as likely cluster members based on a significantly discrepant single-epoch measurement of $v_{\mathrm{los}}$. We calculated the resultant dispersion and repeated the experiment for a large number of random samples and mock data sets. We kept track of the rate of radial velocity variables that would have been detected. For realistic binary fractions and orbital parameter distributions, this should be consistent with the observed rate of variables observed in our sample of cluster members ($\sim 2.5\%$ in the radial region of the two outermost bins in the top panel of Fig.~\ref{fig:dispersion_profile}).

Fig.~\ref{fig:binary_dispersion} shows the probability distribution of the final dispersion for different binary fractions. Different binary fractions result in different measured velocity dispersion differences. With the initial cluster dispersion of $1.5\,\mathrm{km\,s^{-1}}$, the probability of producing a dispersion larger than $\sim 2\,\mathrm{km\,s^{-1}}$ is 1.6\% given a 5\% binary fraction, 23\% given a 10\% binary fraction, and a velocity dispersion of $2\,\mathrm{km\,s^{-1}}$ or more is easily obtained for a binary fraction higher than 20\%. The probability of producing a dispersion larger than $2.5\,\mathrm{km\,s^{-1}}$ (as observed at a projected radius of $\sim1000$ arcsec; Fig.~\ref{fig:dispersion_profile}) is 0\% given a 5\% binary fraction, 0.04\% given a 10\% binary fraction, and 12.4\% given a 20\% binary fraction.  However, \citet{2019A&A...632A...3G} show the core binary fraction in NGC 3201 is $6.75\%\pm0.72\%$, which decreases outwards \citep{2016MNRAS.455.3009M} with radius. In addition, with deep field observations out to 8 arcmin \citep{2018MNRAS.476..271S}, the tight main sequence track argues against a high binary fraction. The observed rate of radial velocity variables (2.5\%) in our sample of cluster members in the two outer radial bins of Fig.~\ref{fig:dispersion_profile} also argues against a binary fraction significantly larger than 10\%. Adopting a binary fraction of 20\% or higher in our mock radial velocity experiments yields a typical rate of detected radial velocity variables in excess of 4\%. A binary fraction large enough to significantly inflate the velocity dispersion would also overpopulate the wings of the velocity distribution compared to the observed sample. For example, with a 20\% binary fraction, we would expect in excess of 15 stars outside the 5$\sigma$ range shown in Fig.~\ref{fig:velocity_radius} at radii beyond $1000\,\mathrm{arcsec}$ even before considering non-members, which is already more than we observe.

Hence, although we cannot exclude that undetected binaries contribute to inflating the observed velocity dispersion, we conclude that the underestimation of the dispersion in the outer parts of the GC is unlikely to be purely due to the effect of undetected binaries given that the binary fraction is likely to be smaller than 10\%.

\subsection{The effect of different escape rates}

As we can see from  Fig.~\ref{fig:GC_orbit}, interacting with the MW  produces the tidal tails from the escaped stars. The unbound stars in the tidal tails might increase the measured LOS velocity dispersion depending on the viewing angle. The escape rate, which describes the efficiency with which stars escape from the GC, will determine the number of stars inside the tidal tails, and thus might change the dispersion. 

Stellar-mass BHs are believed to be able to shape the core profiles of globular clusters and increase the escape rate of stars. The GCs in the MW possess a clear separation in the distribution of core radii into `core collapsed' and `non core collapsed' clusters, defined by small and large core radii, respectively. With strong gravitational interaction, the BHs effectively deposit energy into the GC bulk population, leading to a `puffier' core \citep[e.g.][]{2004ApJ...608L..25M,2007MNRAS.379L..40M,2008MNRAS.386...65M,2016MNRAS.462.2333P}. In the outer parts, BHs can increase the escape rate of the cluster by close interaction with other stars \citep{2019MNRAS.487.2412G,2020MNRAS.491.2413W}, resulting in a larger number of stars in the tidal tails. NGC~3201 is known to host stellar mass BHs from radial velocity measurements \citep{2018MNRAS.475L..15G}, and the luminosity profile of the core region \citep{2018MNRAS.478.1844A,2019ApJ...871...38K}. However, the effect on the LOS dispersion  from the BHs at large radii is unknown.

\begin{figure}
    \centering
    \includegraphics[trim={8mm 8mm 8mm 8mm},width=\columnwidth]{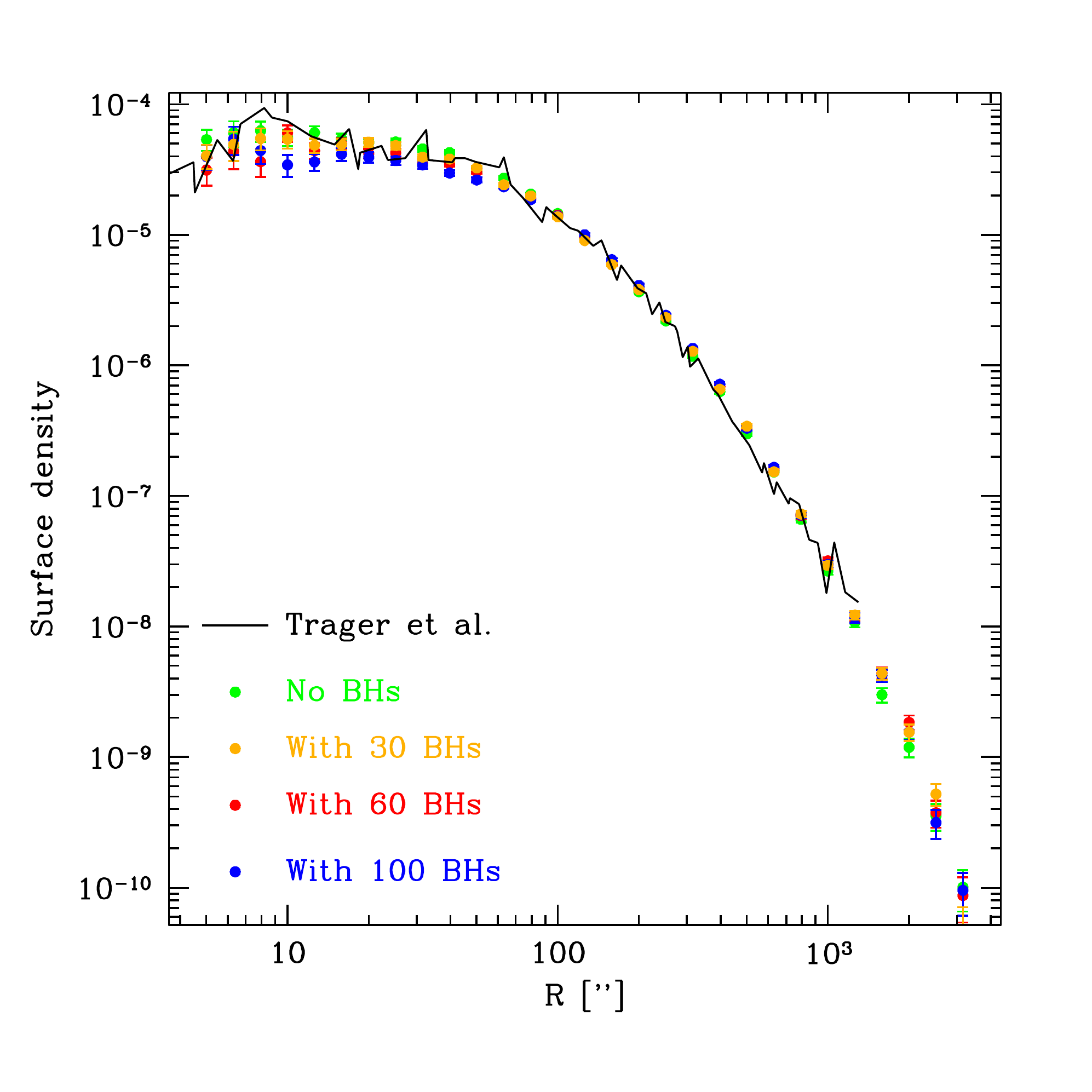}
    \caption{The comparison of the surface density from the simulations with different numbers of BHs. The inner part of the surface density decreases with the number of BHs. However, the effects are insignificant as the GCs with 30 and 60 BHs respectively still fit the data reasonably well.}
    \label{fig:comp_sdens}
\end{figure}

To explore the effects from BHs, we included extra BHs in the $N$-body simulation, but kept all the other parameters the same as the best-fitting model. Following 4 Gyr of evolution of the simulations with varying BH numbers, the 5\% Lagrangian radius (the radius which contains 5\% of the bound mass of the cluster) of the GC differs significantly, whereby the core of the GC with 100 BHs is 3\% larger than that of the GC without BHs. Correspondingly, the surface density profile in the inner part of the GC with BHs is slightly lower than the GC without BHs. However,  Fig.~\ref{fig:comp_sdens} shows that the GCs with 30 and 60 BHs still fit the data reasonably well within the core region. 

\begin{figure}
    \centering
    \includegraphics[width=\columnwidth]{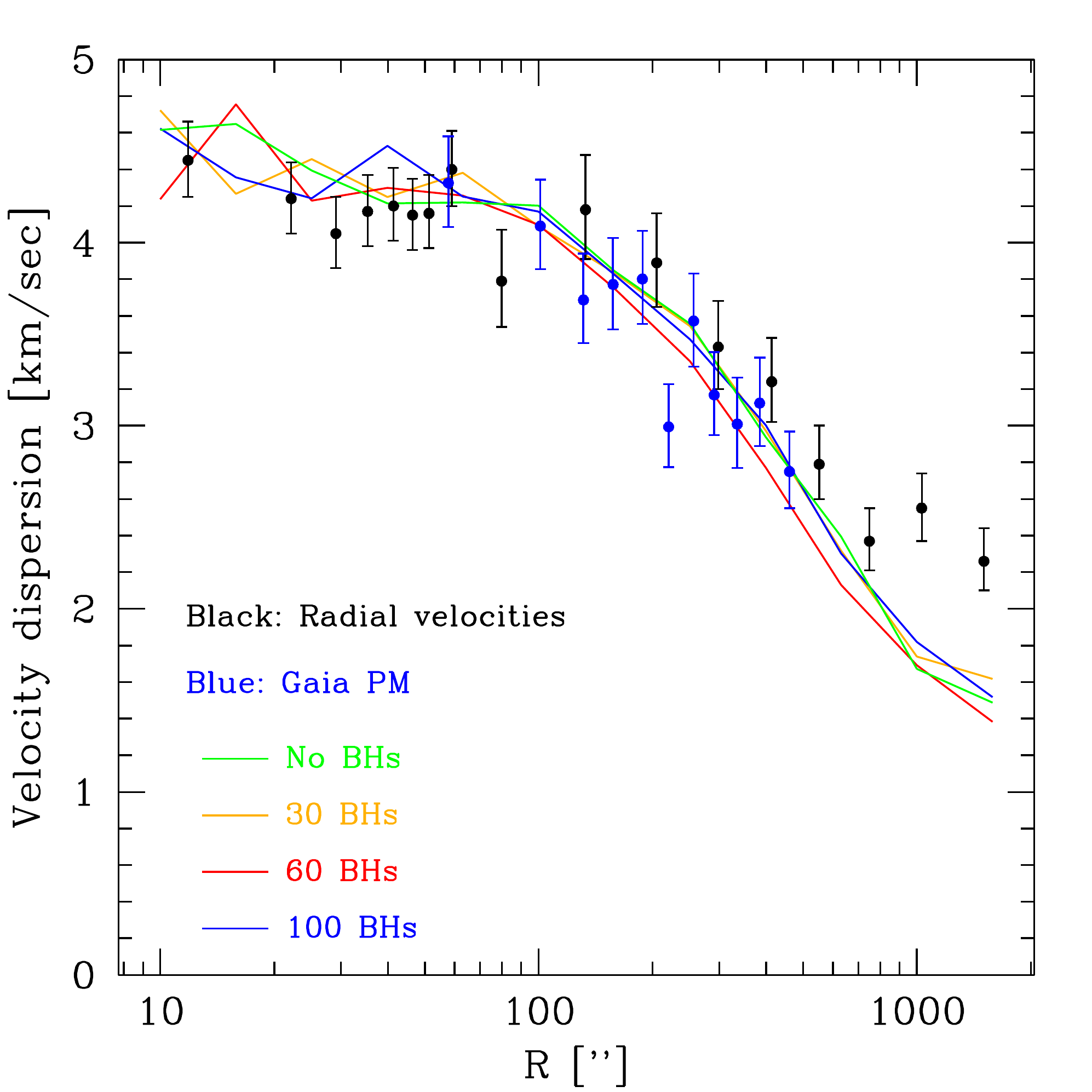}
    \caption{The dispersion profiles of simulated clusters with varying numbers of stellar-mass BHs. Compared to a cluster without BHs, the effects of BHs on the dispersion profile are small. None of the simulated GCs are able to reproduce the large observed velocity dispersion in the outer cluster region.}
    \label{fig:comp_disp_bh}
\end{figure}

Compared to the cluster without BHs, the escape rate is about 1.3\% higher for the cluster with 30 BHs, and is about 2.6\% higher for the cluster with 60 BHs. However, the effects on the dispersion profile are insignificant. As  Fig.~\ref{fig:comp_disp_bh} shows, the clusters with 30 and 60 BHs have slightly lower dispersion, whereas the cluster with 100 BHs has a higher dispersion, suggesting that the effects from BHs are less significant than the systematic uncertainties on the dispersion profile. In the more extreme simulation with 150 BHs, we found that the cluster core is strongly heated, where the 5\% Lagrangian radius is about 0.22 pc after 4 Gyr of evolution. However, the escape rate is not significantly different to the other simulations. Hence we conclude that BHs are not able to produce the observed dispersion.

We also estimated the effect of a large escape rate directly with a mock tidal tail model. We adopt a Lagrange Stripping technique \citep{2015ApJ...803...80K,  2015MNRAS.452..301F} to produce an oversampled tidal tail model. The progenitor was assumed to have a mass of $1.5\times10^{5}\,\mathrm{M_{\odot}}$ and used the same initial conditions as the $N$-body model. The tails were evolved for 200 Myr, releasing particles every 0.005 Myr in the MWPotential2014 \citep{2015ApJS..216...29B}, which is nearly identical to the Irrgang potential in the inner MW and matches the $N$-body simulations quite well. The simulation was performed using the dynamics package \texttt{gala} \citep{gala}. However, we found that with a significantly larger escape rate, the final dispersion is still about $1.5\,\mathrm{km\,s^{-1}}$, which is roughly consistent with the $N$-body model, but still lower than the observations. In the tail model, we find 405 tail stars projected within $r_{\rm Jacobi}$. Because the escape rate in this model is a factor of 10-30 too high, and only a small fraction of the escaping stars are bright enough for our observations, we conclude that tail stars have a negligible contribution to the kinematics within $r_{\rm Jacobi}$.

\section{Conclusions}
\label{sec:conclusions}
In this paper, we described our GC survey with 2dF/AAOmega and the first results for the GC NGC~3201. Aiming at constraining their evolutionary history, we acquired spectra of stars in the outer part of five GCs (see Tab.~\ref{tab:observation_log}). Those observations are designed to be divided into different blocks separated by at least one month so that we could detect short-period binaries. 

We select stars near the turn off, sub-giant branch and in the RGB phase as our targets, as those stars have three strong calcium absorption lines (CaT) at near-infrared wavelengths. Templates based on the three lines are fitted to the spectrum to extract information from our observations, from which we can determine the redshift and stellar parameters, and hence the LOS velocity and stellar properties like metallicity. The detailed study and comparison of the five GCs will be presented in future work. 

As the first result of the survey, we discussed the dynamics of NGC~3201 from our observations. A dark matter free $N$-body simulation, that includes the effect from the MW potential, is built and compared with the observations. We confirm the LOS velocity gradient observed in the GC comes mainly from perspective rotation effects. In addition, we found a weak rotational signal in the inner part of the GC with amplitude of $\sim 1\,\mathrm{km\,s^{-1}}$. The PA of this signal is different to the tidal tails and the potential escapers, which suggests that it comes from the internal rotation of the GC. Besides, we found a rotational signal at the outer part of the GC that has the same rotational direction compared to the tidal tails and potential escapers. However, within the field of view, the contribution from tidal tails are limited,  suggesting that the stars at the outskirt are likely potential escapers.

We also discussed the dispersion profile of the GC. Compared to the simulation, the observed dispersion profile is lower beyond the King tidal radius. We discussed the potential source of this discrepancy. Effects due to the interaction of the cluster with the MW potential and potential escapers are included in the $N$-body model,, hence we conclude that both the MW potential and potential escapers cannot solve the discrepancy. With the simulations that include BHs, we found adding BHs in simulation can increase the escape rate of the cluster, but the change in dispersion is insignificant. Mock tidal tails produced with large escape rates also have a small dispersion compared to the data, consistent with the $N$-body model. In addition, we performed an analysis of the effect of binaries with multi-epoch observations, finding that they should be taken into account, but are unlikely to fully explain the difference in dispersion.

Since the dispersion relates to the dynamical mass, the presence of dark matter \citep[e.g.][]{1984ApJ...277..470P,2015ApJ...808L..35T} at larger radii would naturally lead to a flattened dispersion. \citet{2019ApJ...887L..12B} discuss the possibility that NGC~3201 is embedded in a dark matter halo. Although this can naturally explain an increased dispersion, there is still no direct evidence for the existence of dark matter in globular clusters. In addition, the signal of unbounded stars at the outer part of the GC, as well as the evidence of tidal streams associated with NGC~3201 \citep[][]{2010ApJ...721.1790C, 2014A&A...572A..30K, 2016MNRAS.457.2078A,2020arXiv201014381P, 2020arXiv201205245I}, argue against the presence of dark matter. Hence the existence of the dark matter needs further confirmation. One could argue that we see the final phases of the stripping of the dark matter halo, e.g. if NGC~3201 was in the nuclear cluster of the dwarf galaxy, which could explain why the stars are also affected by tides. However, NGC~3201 is association with the Gaia-Enceladus/Sequoia accretion \citep[][]{2019MNRAS.488.1235M, 2019A&A...630L...4M}, which suggests that this cluster was accreted $\gtrsim9$ Gyr ago as part of a dwarf galaxy with multiple star clusters, among which $\omega$ Centauri. The high mass, and multiple metallicities of $\omega$ Centauri make this cluster a much more plausible `former nuclear cluster' compared to NGC~3201. We note that there are some additional effects that can potentially influence the dispersion. The orbital phase of this cluster is less constrained due the uncertainties in proper motion, distance and the galactic potential, even though the position of the NGC~3201 on the sky is well known. Also, an initially rotating cluster in the simulation might result in a different dispersion profile. Finally, interactions with sub-structure in the MW -- either baryonic or non-baryonic -- may have heated the stars in the cluster and in the tails \citep[e.g.][]{2017MNRAS.470...60E}. This discrepancy suggests that there is more we can learn from the dynamics of the outer part of the GC on its evolutionary history. Analyses of the dark matter content/distribution, as well as the unexplored effects on the velocity gradient, dispersion and tidal tails, will be presented in our future work.

\section*{Acknowledgements}
ZW is supported by a Dean's International Postgraduate Research Scholarship at the University of Sydney. WHO gratefully acknowledges financial support through the Hunstead Student Support Scholarship from the Dick Hunstead Fund in the University of Sydney's School of Physics. MG, TdB and EB acknowledge financial support from the European Research Council (ERC StG-335936, CLUSTERS). MG acknowledges support from the Ministry of Science and Innovation through a Europa Excelencia grant (EUR2020-112157). VHB acknowledges the support of the Natural Sciences and Engineering Research Council of Canada (NSERC) through grant RGPIN-2020-05990. EB acknowledges financial support from a Vici grant from the Netherlands Organisation for Scientific Research (NWO). We thank Paolo Bianchini for providing the dispersion profile from his proper motion studies.
Based in part on data acquired through the Australian Astronomical Observatory. We acknowledge the traditional owners of the land on which the AAT stands, the Gamilaraay people, and pay our respects to elders past, present and emerging.

Parts of this work were performed on the OzSTAR national facility at Swinburne University of Technology. The OzSTAR program receives funding in part from the National Collaborative Research Infrastructure Strategy (NCRIS) Astronomy allocation provided by the Australian Government.
 
\section*{Data Availability}
The data underlying this article may be made available on reasonable request to the corresponding author.




\bibliographystyle{mnras}
\bibliography{ref} 

\begin{thebibliography}{}
\makeatletter
\relax
\def\mn@urlcharsother{\let\do\@makeother \do\$\do\&\do\#\do\^\do\_\do\%\do\~}
\def\mn@doi{\begingroup\mn@urlcharsother \@ifnextchar [ {\mn@doi@}
  {\mn@doi@[]}}
\def\mn@doi@[#1]#2{\def\@tempa{#1}\ifx\@tempa\@empty \href
  {http://dx.doi.org/#2} {doi:#2}\else \href {http://dx.doi.org/#2} {#1}\fi
  \endgroup}
\def\mn@eprint#1#2{\mn@eprint@#1:#2::\@nil}
\def\mn@eprint@arXiv#1{\href {http://arxiv.org/abs/#1} {{\tt arXiv:#1}}}
\def\mn@eprint@dblp#1{\href {http://dblp.uni-trier.de/rec/bibtex/#1.xml}
  {dblp:#1}}
\def\mn@eprint@#1:#2:#3:#4\@nil{\def\@tempa {#1}\def\@tempb {#2}\def\@tempc
  {#3}\ifx \@tempc \@empty \let \@tempc \@tempb \let \@tempb \@tempa \fi \ifx
  \@tempb \@empty \def\@tempb {arXiv}\fi \@ifundefined
  {mn@eprint@\@tempb}{\@tempb:\@tempc}{\expandafter \expandafter \csname
  mn@eprint@\@tempb\endcsname \expandafter{\@tempc}}}

\bibitem[\protect\citeauthoryear{{AAO Software Team}}{{AAO Software
  Team}}{2015}]{2015ascl.soft05015A}
{AAO Software Team} 2015, {2dfdr: Data reduction software} (\mn@eprint {ascl}
  {1505.015})

\bibitem[\protect\citeauthoryear{{Anguiano} et~al.,}{{Anguiano}
  et~al.}{2016}]{2016MNRAS.457.2078A}
{Anguiano} B.,  et~al., 2016, \mn@doi [\mnras] {10.1093/mnras/stw083}, \href
  {https://ui.adsabs.harvard.edu/abs/2016MNRAS.457.2078A} {457, 2078}

\bibitem[\protect\citeauthoryear{{Askar}, {Arca Sedda}  \& {Giersz}}{{Askar}
  et~al.}{2018}]{2018MNRAS.478.1844A}
{Askar} A.,  {Arca Sedda} M.,   {Giersz} M.,  2018, \mn@doi [\mnras]
  {10.1093/mnras/sty1186}, \href
  {http://adsabs.harvard.edu/abs/2018MNRAS.478.1844A} {478, 1844}

\bibitem[\protect\citeauthoryear{{Balbinot} \& {Gieles}}{{Balbinot} \&
  {Gieles}}{2018}]{2018MNRAS.474.2479B}
{Balbinot} E.,  {Gieles} M.,  2018, \mn@doi [\mnras] {10.1093/mnras/stx2708},
  \href {https://ui.adsabs.harvard.edu/abs/2018MNRAS.474.2479B} {474, 2479}

\bibitem[\protect\citeauthoryear{{Baumgardt}}{{Baumgardt}}{2001}]{2001MNRAS.325.1323B}
{Baumgardt} H.,  2001, \mn@doi [\mnras] {10.1046/j.1365-8711.2001.04272.x},
  \href
  {http://adsabs.harvard.edu/cgi-bin/nph-bib_query?bibcode=2001MNRAS.325.1323B&db_key=AST}
  {325, 1323}

\bibitem[\protect\citeauthoryear{{Baumgardt}}{{Baumgardt}}{2017}]{2017MNRAS.464.2174B}
{Baumgardt} H.,  2017, \mn@doi [\mnras] {10.1093/mnras/stw2488}, \href
  {https://ui.adsabs.harvard.edu/abs/2017MNRAS.464.2174B} {464, 2174}

\bibitem[\protect\citeauthoryear{{Baumgardt} \& {Hilker}}{{Baumgardt} \&
  {Hilker}}{2018}]{2018MNRAS.478.1520B}
{Baumgardt} H.,  {Hilker} M.,  2018, \mn@doi [\mnras] {10.1093/mnras/sty1057},
  \href {https://ui.adsabs.harvard.edu/abs/2018MNRAS.478.1520B} {478, 1520}

\bibitem[\protect\citeauthoryear{{Baumgardt} \& {Makino}}{{Baumgardt} \&
  {Makino}}{2003}]{2003MNRAS.340..227B}
{Baumgardt} H.,  {Makino} J.,  2003, \mn@doi [\mnras]
  {10.1046/j.1365-8711.2003.06286.x}, \href
  {http://adsabs.harvard.edu/cgi-bin/nph-bib_query?bibcode=2003MNRAS.340..227B&db_key=AST}
  {340, 227}

\bibitem[\protect\citeauthoryear{{Baumgardt} \& {Mieske}}{{Baumgardt} \&
  {Mieske}}{2008}]{2008MNRAS.391..942B}
{Baumgardt} H.,  {Mieske} S.,  2008, \mn@doi [\mnras]
  {10.1111/j.1365-2966.2008.13949.x}, \href
  {https://ui.adsabs.harvard.edu/abs/2008MNRAS.391..942B} {391, 942}

\bibitem[\protect\citeauthoryear{{Baumgardt}, {Hilker}, {Sollima}  \&
  {Bellini}}{{Baumgardt} et~al.}{2019}]{baumgardtetal2019}
{Baumgardt} H.,  {Hilker} M.,  {Sollima} A.,   {Bellini} A.,  2019, \mn@doi
  [\mnras] {10.1093/mnras/sty2997}, \href
  {https://ui.adsabs.harvard.edu/abs/2019MNRAS.482.5138B} {482, 5138}

\bibitem[\protect\citeauthoryear{{Bellazzini}, {Bragaglia}, {Carretta},
  {Gratton}, {Lucatello}, {Catanzaro}  \& {Leone}}{{Bellazzini}
  et~al.}{2012}]{2012A&A...538A..18B}
{Bellazzini} M.,  {Bragaglia} A.,  {Carretta} E.,  {Gratton} R.~G.,
  {Lucatello} S.,  {Catanzaro} G.,   {Leone} F.,  2012, \mn@doi [\aap]
  {10.1051/0004-6361/201118056}, \href
  {https://ui.adsabs.harvard.edu/abs/2012A&A...538A..18B} {538, A18}

\bibitem[\protect\citeauthoryear{{Bianchini}, {van der Marel}, {del Pino},
  {Watkins}, {Bellini}, {Fardal}, {Libralato}  \& {Sills}}{{Bianchini}
  et~al.}{2018}]{2018MNRAS.481.2125B}
{Bianchini} P.,  {van der Marel} R.~P.,  {del Pino} A.,  {Watkins} L.~L.,
  {Bellini} A.,  {Fardal} M.~A.,  {Libralato} M.,   {Sills} A.,  2018, \mn@doi
  [\mnras] {10.1093/mnras/sty2365}, \href
  {https://ui.adsabs.harvard.edu/abs/2018MNRAS.481.2125B} {481, 2125}

\bibitem[\protect\citeauthoryear{{Bianchini}, {Ibata}  \& {Famaey}}{{Bianchini}
  et~al.}{2019}]{2019ApJ...887L..12B}
{Bianchini} P.,  {Ibata} R.,   {Famaey} B.,  2019, \mn@doi [\apjl]
  {10.3847/2041-8213/ab58d1}, \href
  {https://ui.adsabs.harvard.edu/abs/2019ApJ...887L..12B} {887, L12}

\bibitem[\protect\citeauthoryear{{Bovy}}{{Bovy}}{2015}]{2015ApJS..216...29B}
{Bovy} J.,  2015, \mn@doi [\apjs] {10.1088/0067-0049/216/2/29}, \href
  {https://ui.adsabs.harvard.edu/abs/2015ApJS..216...29B} {216, 29}

\bibitem[\protect\citeauthoryear{{Cai}, {Gieles}, {Heggie}  \& {Varri}}{{Cai}
  et~al.}{2016}]{2016MNRAS.455..596C}
{Cai} M.~X.,  {Gieles} M.,  {Heggie} D.~C.,   {Varri} A.~L.,  2016, \mn@doi
  [\mnras] {10.1093/mnras/stv2325}, \href
  {http://adsabs.harvard.edu/abs/2016MNRAS.455..596C} {455, 596}

\bibitem[\protect\citeauthoryear{{Carballo-Bello}, {Gieles}, {Sollima},
  {Koposov}, {Mart{\'\i}nez-Delgado}  \& {Pe{\~n}arrubia}}{{Carballo-Bello}
  et~al.}{2012}]{2012MNRAS.419...14C}
{Carballo-Bello} J.~A.,  {Gieles} M.,  {Sollima} A.,  {Koposov} S.,
  {Mart{\'\i}nez-Delgado} D.,   {Pe{\~n}arrubia} J.,  2012, \mn@doi [\mnras]
  {10.1111/j.1365-2966.2011.19663.x}, \href
  {https://ui.adsabs.harvard.edu/abs/2012MNRAS.419...14C} {419, 14}

\bibitem[\protect\citeauthoryear{{Chen} \& {Chen}}{{Chen} \&
  {Chen}}{2010}]{2010ApJ...721.1790C}
{Chen} C.~W.,  {Chen} W.~P.,  2010, \mn@doi [\apj]
  {10.1088/0004-637X/721/2/1790}, \href
  {https://ui.adsabs.harvard.edu/abs/2010ApJ...721.1790C} {721, 1790}

\bibitem[\protect\citeauthoryear{{Chun} et~al.,}{{Chun}
  et~al.}{2010}]{2010AJ....139..606C}
{Chun} S.-H.,  et~al., 2010, \mn@doi [\aj] {10.1088/0004-6256/139/2/606}, \href
  {https://ui.adsabs.harvard.edu/abs/2010AJ....139..606C} {139, 606}

\bibitem[\protect\citeauthoryear{{Claydon}, {Gieles}  \& {Zocchi}}{{Claydon}
  et~al.}{2017}]{2017MNRAS.466.3937C}
{Claydon} I.,  {Gieles} M.,   {Zocchi} A.,  2017, \mn@doi [\mnras]
  {10.1093/mnras/stw3309}, \href
  {https://ui.adsabs.harvard.edu/abs/2017MNRAS.466.3937C} {466, 3937}

\bibitem[\protect\citeauthoryear{{Claydon}, {Gieles}, {Varri}, {Heggie}  \&
  {Zocchi}}{{Claydon} et~al.}{2019}]{2019MNRAS.487..147C}
{Claydon} I.,  {Gieles} M.,  {Varri} A.~L.,  {Heggie} D.~C.,   {Zocchi} A.,
  2019, \mn@doi [\mnras] {10.1093/mnras/stz1109}, \href
  {https://ui.adsabs.harvard.edu/abs/2019MNRAS.487..147C} {487, 147}

\bibitem[\protect\citeauthoryear{{Conroy}, {Loeb}  \& {Spergel}}{{Conroy}
  et~al.}{2011}]{2011ApJ...741...72C}
{Conroy} C.,  {Loeb} A.,   {Spergel} D.~N.,  2011, \mn@doi [\apj]
  {10.1088/0004-637X/741/2/72}, \href
  {https://ui.adsabs.harvard.edu/abs/2011ApJ...741...72C} {741, 72}

\bibitem[\protect\citeauthoryear{{Cote}, {Welch}, {Fischer}  \&
  {Gebhardt}}{{Cote} et~al.}{1995}]{1995ApJ...454..788C}
{Cote} P.,  {Welch} D.~L.,  {Fischer} P.,   {Gebhardt} K.,  1995, \mn@doi
  [\apj] {10.1086/176532}, \href
  {https://ui.adsabs.harvard.edu/abs/1995ApJ...454..788C} {454, 788}

\bibitem[\protect\citeauthoryear{{Cottaar} \& {H{\'e}nault-Brunet}}{{Cottaar}
  \& {H{\'e}nault-Brunet}}{2014}]{2014A&A...562A..20C}
{Cottaar} M.,  {H{\'e}nault-Brunet} V.,  2014, \mn@doi [\aap]
  {10.1051/0004-6361/201322445}, \href
  {https://ui.adsabs.harvard.edu/abs/2014A&A...562A..20C} {562, A20}

\bibitem[\protect\citeauthoryear{{Cottaar}, {Meyer}  \& {Parker}}{{Cottaar}
  et~al.}{2012}]{2012A&A...547A..35C}
{Cottaar} M.,  {Meyer} M.~R.,   {Parker} R.~J.,  2012, \mn@doi [\aap]
  {10.1051/0004-6361/201219673}, \href
  {https://ui.adsabs.harvard.edu/abs/2012A&A...547A..35C} {547, A35}

\bibitem[\protect\citeauthoryear{{Da Costa}}{{Da
  Costa}}{2012}]{2012ApJ...751....6D}
{Da Costa} G.~S.,  2012, \mn@doi [\apj] {10.1088/0004-637X/751/1/6}, \href
  {https://ui.adsabs.harvard.edu/abs/2012ApJ...751....6D} {751, 6}

\bibitem[\protect\citeauthoryear{{Daniel}, {Heggie}  \& {Varri}}{{Daniel}
  et~al.}{2017}]{2017MNRAS.468.1453D}
{Daniel} K.~J.,  {Heggie} D.~C.,   {Varri} A.~L.,  2017, \mn@doi [\mnras]
  {10.1093/mnras/stx571}, \href
  {https://ui.adsabs.harvard.edu/abs/2017MNRAS.468.1453D} {468, 1453}

\bibitem[\protect\citeauthoryear{Edl{\'e}n \& Risberg}{Edl{\'e}n \&
  Risberg}{1956}]{edlen1956spectrum}
Edl{\'e}n B.,  Risberg P.,  1956, Arkiv For Fysik, 10, 553

\bibitem[\protect\citeauthoryear{{Erkal}, {Koposov}  \& {Belokurov}}{{Erkal}
  et~al.}{2017}]{2017MNRAS.470...60E}
{Erkal} D.,  {Koposov} S.~E.,   {Belokurov} V.,  2017, \mn@doi [\mnras]
  {10.1093/mnras/stx1208}, \href
  {https://ui.adsabs.harvard.edu/abs/2017MNRAS.470...60E} {470, 60}

\bibitem[\protect\citeauthoryear{{Fardal}, {Huang}  \& {Weinberg}}{{Fardal}
  et~al.}{2015}]{2015MNRAS.452..301F}
{Fardal} M.~A.,  {Huang} S.,   {Weinberg} M.~D.,  2015, \mn@doi [\mnras]
  {10.1093/mnras/stv1198}, \href
  {https://ui.adsabs.harvard.edu/abs/2015MNRAS.452..301F} {452, 301}

\bibitem[\protect\citeauthoryear{{Ferraro} et~al.,}{{Ferraro}
  et~al.}{2018}]{2018ApJ...860...50F}
{Ferraro} F.~R.,  et~al., 2018, \mn@doi [\apj] {10.3847/1538-4357/aabe2f},
  \href {https://ui.adsabs.harvard.edu/abs/2018ApJ...860...50F} {860, 50}

\bibitem[\protect\citeauthoryear{{Foreman-Mackey}, {Hogg}, {Lang}  \&
  {Goodman}}{{Foreman-Mackey} et~al.}{2013}]{2013PASP..125..306F}
{Foreman-Mackey} D.,  {Hogg} D.~W.,  {Lang} D.,   {Goodman} J.,  2013, \mn@doi
  [\pasp] {10.1086/670067}, \href
  {https://ui.adsabs.harvard.edu/abs/2013PASP..125..306F} {125, 306}

\bibitem[\protect\citeauthoryear{{Fukushige} \& {Heggie}}{{Fukushige} \&
  {Heggie}}{2000}]{2000MNRAS.318..753F}
{Fukushige} T.,  {Heggie} D.~C.,  2000, \mnras, \href
  {http://adsabs.harvard.edu/cgi-bin/nph-bib_query?bibcode=2000MNRAS.318..753F&db_key=AST}
  {318, 753}

\bibitem[\protect\citeauthoryear{{Gaia Collaboration} et~al.,}{{Gaia
  Collaboration} et~al.}{2018a}]{2018A&A...616A...1G}
{Gaia Collaboration} et~al., 2018a, \mn@doi [\aap]
  {10.1051/0004-6361/201833051}, \href
  {https://ui.adsabs.harvard.edu/abs/2018A&A...616A...1G} {616, A1}

\bibitem[\protect\citeauthoryear{{Gaia Collaboration} et~al.,}{{Gaia
  Collaboration} et~al.}{2018b}]{2018A&A...616A..10G}
{Gaia Collaboration} et~al., 2018b, \mn@doi [\aap]
  {10.1051/0004-6361/201832843}, \href
  {https://ui.adsabs.harvard.edu/abs/2018A&A...616A..10G} {616, A10}

\bibitem[\protect\citeauthoryear{{Gaia Collaboration} et~al.,}{{Gaia
  Collaboration} et~al.}{2018c}]{2018A&A...616A..12G}
{Gaia Collaboration} et~al., 2018c, \mn@doi [\aap]
  {10.1051/0004-6361/201832698}, \href
  {https://ui.adsabs.harvard.edu/abs/2018A&A...616A..12G} {616, A12}

\bibitem[\protect\citeauthoryear{{Gieles} \& {Zocchi}}{{Gieles} \&
  {Zocchi}}{2015}]{2015MNRAS.454..576G}
{Gieles} M.,  {Zocchi} A.,  2015, \mn@doi [\mnras] {10.1093/mnras/stv1848},
  \href {http://adsabs.harvard.edu/abs/2015MNRAS.454..576G} {454, 576}

\bibitem[\protect\citeauthoryear{{Gieles}, {Balbinot}, {Yaaqib},
  {H{\'e}nault-Brunet}, {Zocchi}, {Peuten}  \& {Jonker}}{{Gieles}
  et~al.}{2018}]{2018MNRAS.473.4832G}
{Gieles} M.,  {Balbinot} E.,  {Yaaqib} R. I.~S.~M.,  {H{\'e}nault-Brunet} V.,
  {Zocchi} A.,  {Peuten} M.,   {Jonker} P.~G.,  2018, \mn@doi [\mnras]
  {10.1093/mnras/stx2694}, \href
  {https://ui.adsabs.harvard.edu/abs/2018MNRAS.473.4832G} {473, 4832}

\bibitem[\protect\citeauthoryear{{Giersz}, {Askar}, {Wang}, {Hypki}, {Leveque}
  \& {Spurzem}}{{Giersz} et~al.}{2019}]{2019MNRAS.487.2412G}
{Giersz} M.,  {Askar} A.,  {Wang} L.,  {Hypki} A.,  {Leveque} A.,   {Spurzem}
  R.,  2019, \mn@doi [\mnras] {10.1093/mnras/stz1460}, \href
  {https://ui.adsabs.harvard.edu/abs/2019MNRAS.487.2412G} {487, 2412}

\bibitem[\protect\citeauthoryear{{Giesers} et~al.,}{{Giesers}
  et~al.}{2018}]{2018MNRAS.475L..15G}
{Giesers} B.,  et~al., 2018, \mn@doi [\mnras] {10.1093/mnrasl/slx203}, \href
  {https://ui.adsabs.harvard.edu/abs/2018MNRAS.475L..15G} {475, L15}

\bibitem[\protect\citeauthoryear{{Giesers} et~al.,}{{Giesers}
  et~al.}{2019}]{2019A&A...632A...3G}
{Giesers} B.,  et~al., 2019, \mn@doi [\aap] {10.1051/0004-6361/201936203},
  \href {https://ui.adsabs.harvard.edu/abs/2019A&A...632A...3G} {632, A3}

\bibitem[\protect\citeauthoryear{{Hansen}, {Riley}, {Strigari}, {Marshall},
  {Ferguson}, {Zepeda}  \& {Sneden}}{{Hansen}
  et~al.}{2020}]{2020arXiv200712165H}
{Hansen} T.~T.,  {Riley} A.~H.,  {Strigari} L.~E.,  {Marshall} J.~L.,
  {Ferguson} P.~S.,  {Zepeda} J.,   {Sneden} C.,  2020, arXiv e-prints, \href
  {https://ui.adsabs.harvard.edu/abs/2020arXiv200712165H} {p. arXiv:2007.12165}

\bibitem[\protect\citeauthoryear{{Harris}}{{Harris}}{1996}]{1996AJ....112.1487H}
{Harris} W.~E.,  1996, \mn@doi [\aj] {10.1086/118116}, \href
  {https://ui.adsabs.harvard.edu/abs/1996AJ....112.1487H} {112, 1487}

\bibitem[\protect\citeauthoryear{{Harris}}{{Harris}}{2010}]{Harris10}
{Harris} W.~E.,  2010, preprint, \href
  {http://adsabs.harvard.edu/abs/2010arXiv1012.3224H} {} (\mn@eprint {arXiv}
  {1012.3224})

\bibitem[\protect\citeauthoryear{{Helmi} et~al.,}{{Helmi}
  et~al.}{2018}]{Helmi18}
{Helmi} A.,  et~al., 2018, \mn@doi [\aap] {10.1051/0004-6361/201832698}, \href
  {https://ui.adsabs.harvard.edu/#abs/2018A&A...616A..12G} {616, A12}

\bibitem[\protect\citeauthoryear{{H{\'e}nault-Brunet}, {Gieles}, {Sollima},
  {Watkins}, {Zocchi}, {Claydon}, {Pancino}  \&
  {Baumgardt}}{{H{\'e}nault-Brunet} et~al.}{2019}]{2019MNRAS.483.1400H}
{H{\'e}nault-Brunet} V.,  {Gieles} M.,  {Sollima} A.,  {Watkins} L.~L.,
  {Zocchi} A.,  {Claydon} I.,  {Pancino} E.,   {Baumgardt} H.,  2019, \mn@doi
  [\mnras] {10.1093/mnras/sty3187}, \href
  {http://adsabs.harvard.edu/abs/2019MNRAS.483.1400H} {483, 1400}

\bibitem[\protect\citeauthoryear{{Henon}}{{Henon}}{1970}]{1970A&A.....9...24H}
{Henon} M.,  1970, \aap, \href
  {https://ui.adsabs.harvard.edu/abs/1970A&A.....9...24H} {9, 24}

\bibitem[\protect\citeauthoryear{{Ibata} et~al.,}{{Ibata}
  et~al.}{2020}]{2020arXiv201205245I}
{Ibata} R.,  et~al., 2020, arXiv e-prints, \href
  {https://ui.adsabs.harvard.edu/abs/2020arXiv201205245I} {p. arXiv:2012.05245}

\bibitem[\protect\citeauthoryear{{Irrgang}, {Wilcox}, {Tucker}  \&
  {Schiefelbein}}{{Irrgang} et~al.}{2013}]{Irrgangetal2013}
{Irrgang} A.,  {Wilcox} B.,  {Tucker} E.,   {Schiefelbein} L.,  2013, \mn@doi
  [\aap] {10.1051/0004-6361/201220540}, \href
  {https://ui.adsabs.harvard.edu/abs/2013A&A...549A.137I} {549, A137}

\bibitem[\protect\citeauthoryear{{Jordi} \& {Grebel}}{{Jordi} \&
  {Grebel}}{2010}]{2010A&A...522A..71J}
{Jordi} K.,  {Grebel} E.~K.,  2010, \mn@doi [\aap]
  {10.1051/0004-6361/201014392}, \href
  {https://ui.adsabs.harvard.edu/abs/2010A&A...522A..71J} {522, A71}

\bibitem[\protect\citeauthoryear{{Kamann} et~al.,}{{Kamann}
  et~al.}{2018}]{2018MNRAS.473.5591K}
{Kamann} S.,  et~al., 2018, \mn@doi [\mnras] {10.1093/mnras/stx2719}, \href
  {https://ui.adsabs.harvard.edu/abs/2018MNRAS.473.5591K} {473, 5591}

\bibitem[\protect\citeauthoryear{{Kimmig}, {Seth}, {Ivans}, {Strader},
  {Caldwell}, {Anderton}  \& {Gregersen}}{{Kimmig}
  et~al.}{2015}]{2015AJ....149...53K}
{Kimmig} B.,  {Seth} A.,  {Ivans} I.~I.,  {Strader} J.,  {Caldwell} N.,
  {Anderton} T.,   {Gregersen} D.,  2015, \mn@doi [\aj]
  {10.1088/0004-6256/149/2/53}, \href
  {https://ui.adsabs.harvard.edu/abs/2015AJ....149...53K} {149, 53}

\bibitem[\protect\citeauthoryear{{King}}{{King}}{1962}]{1962AJ.....67..471K}
{King} I.,  1962, \mn@doi [\aj] {10.1086/108756}, \href
  {https://ui.adsabs.harvard.edu/abs/1962AJ.....67..471K} {67, 471}

\bibitem[\protect\citeauthoryear{{King}}{{King}}{1966}]{1966AJ.....71...64K}
{King} I.~R.,  1966, \aj, \href
  {http://adsabs.harvard.edu/cgi-bin/nph-bib_query?bibcode=1966AJ.....71...64K&db_key=AST}
  {71, 64}

\bibitem[\protect\citeauthoryear{{Kremer}, {Chatterjee}, {Ye}, {Rodriguez}  \&
  {Rasio}}{{Kremer} et~al.}{2019}]{2019ApJ...871...38K}
{Kremer} K.,  {Chatterjee} S.,  {Ye} C.~S.,  {Rodriguez} C.~L.,   {Rasio}
  F.~A.,  2019, \mn@doi [\apj] {10.3847/1538-4357/aaf646}, \href
  {https://ui.adsabs.harvard.edu/abs/2019ApJ...871...38K} {871, 38}

\bibitem[\protect\citeauthoryear{{Kroupa}}{{Kroupa}}{2001}]{2001MNRAS.322..231K}
{Kroupa} P.,  2001, \mn@doi [\mnras] {10.1046/j.1365-8711.2001.04022.x}, \href
  {https://ui.adsabs.harvard.edu/abs/2001MNRAS.322..231K} {322, 231}

\bibitem[\protect\citeauthoryear{{Kunder} et~al.,}{{Kunder}
  et~al.}{2014}]{2014A&A...572A..30K}
{Kunder} A.,  et~al., 2014, \mn@doi [\aap] {10.1051/0004-6361/201424113}, \href
  {https://ui.adsabs.harvard.edu/abs/2014A&A...572A..30K} {572, A30}

\bibitem[\protect\citeauthoryear{{K{\"u}pper}, {Kroupa}, {Baumgardt}  \&
  {Heggie}}{{K{\"u}pper} et~al.}{2010}]{2010MNRAS.407.2241K}
{K{\"u}pper} A. H.~W.,  {Kroupa} P.,  {Baumgardt} H.,   {Heggie} D.~C.,  2010,
  \mn@doi [\mnras] {10.1111/j.1365-2966.2010.17084.x}, \href
  {https://ui.adsabs.harvard.edu/abs/2010MNRAS.407.2241K} {407, 2241}

\bibitem[\protect\citeauthoryear{{K{\"u}pper}, {Balbinot}, {Bonaca},
  {Johnston}, {Hogg}, {Kroupa}  \& {Santiago}}{{K{\"u}pper}
  et~al.}{2015}]{2015ApJ...803...80K}
{K{\"u}pper} A. H.~W.,  {Balbinot} E.,  {Bonaca} A.,  {Johnston} K.~V.,  {Hogg}
  D.~W.,  {Kroupa} P.,   {Santiago} B.~X.,  2015, \mn@doi [\apj]
  {10.1088/0004-637X/803/2/80}, \href
  {https://ui.adsabs.harvard.edu/abs/2015ApJ...803...80K} {803, 80}

\bibitem[\protect\citeauthoryear{{Kuzma}, {Da Costa}  \& {Mackey}}{{Kuzma}
  et~al.}{2018}]{2018MNRAS.473.2881K}
{Kuzma} P.~B.,  {Da Costa} G.~S.,   {Mackey} A.~D.,  2018, \mn@doi [\mnras]
  {10.1093/mnras/stx2353}, \href
  {https://ui.adsabs.harvard.edu/abs/2018MNRAS.473.2881K} {473, 2881}

\bibitem[\protect\citeauthoryear{{Lanzoni} et~al.,}{{Lanzoni}
  et~al.}{2018a}]{2018ApJ...861...16L}
{Lanzoni} B.,  et~al., 2018a, \mn@doi [\apj] {10.3847/1538-4357/aac26a}, \href
  {https://ui.adsabs.harvard.edu/abs/2018ApJ...861...16L} {861, 16}

\bibitem[\protect\citeauthoryear{{Lanzoni} et~al.,}{{Lanzoni}
  et~al.}{2018b}]{2018ApJ...865...11L}
{Lanzoni} B.,  et~al., 2018b, \mn@doi [\apj] {10.3847/1538-4357/aad810}, \href
  {https://ui.adsabs.harvard.edu/abs/2018ApJ...865...11L} {865, 11}

\bibitem[\protect\citeauthoryear{{Mackey}, {Wilkinson}, {Davies}  \&
  {Gilmore}}{{Mackey} et~al.}{2007}]{2007MNRAS.379L..40M}
{Mackey} A.~D.,  {Wilkinson} M.~I.,  {Davies} M.~B.,   {Gilmore} G.~F.,  2007,
  \mn@doi [\mnras] {10.1111/j.1745-3933.2007.00330.x}, \href
  {https://ui.adsabs.harvard.edu/abs/2007MNRAS.379L..40M} {379, L40}

\bibitem[\protect\citeauthoryear{{Mackey}, {Wilkinson}, {Davies}  \&
  {Gilmore}}{{Mackey} et~al.}{2008}]{2008MNRAS.386...65M}
{Mackey} A.~D.,  {Wilkinson} M.~I.,  {Davies} M.~B.,   {Gilmore} G.~F.,  2008,
  \mn@doi [\mnras] {10.1111/j.1365-2966.2008.13052.x}, \href
  {https://ui.adsabs.harvard.edu/abs/2008MNRAS.386...65M} {386, 65}

\bibitem[\protect\citeauthoryear{{Marigo} et~al.,}{{Marigo}
  et~al.}{2017}]{Marigo17}
{Marigo} P.,  et~al., 2017, \mn@doi [\apj] {10.3847/1538-4357/835/1/77}, \href
  {http://adsabs.harvard.edu/abs/2017ApJ...835...77M} {835, 77}

\bibitem[\protect\citeauthoryear{{Mar{\'{\i}}n-Franch}
  et~al.,}{{Mar{\'{\i}}n-Franch} et~al.}{2009}]{Marin-Franch09}
{Mar{\'{\i}}n-Franch} A.,  et~al., 2009, \mn@doi [\apj]
  {10.1088/0004-637X/694/2/1498}, \href
  {http://adsabs.harvard.edu/abs/2009ApJ...694.1498M} {694, 1498}

\bibitem[\protect\citeauthoryear{{Marino} et~al.,}{{Marino}
  et~al.}{2014}]{2014MNRAS.442.3044M}
{Marino} A.~F.,  et~al., 2014, \mn@doi [\mnras] {10.1093/mnras/stu1099}, \href
  {https://ui.adsabs.harvard.edu/abs/2014MNRAS.442.3044M} {442, 3044}

\bibitem[\protect\citeauthoryear{{Mashchenko} \& {Sills}}{{Mashchenko} \&
  {Sills}}{2005a}]{2005ApJ...619..243M}
{Mashchenko} S.,  {Sills} A.,  2005a, \mn@doi [\apj] {10.1086/426132}, \href
  {https://ui.adsabs.harvard.edu/abs/2005ApJ...619..243M} {619, 243}

\bibitem[\protect\citeauthoryear{{Mashchenko} \& {Sills}}{{Mashchenko} \&
  {Sills}}{2005b}]{2005ApJ...619..258M}
{Mashchenko} S.,  {Sills} A.,  2005b, \mn@doi [\apj] {10.1086/426133}, \href
  {https://ui.adsabs.harvard.edu/abs/2005ApJ...619..258M} {619, 258}

\bibitem[\protect\citeauthoryear{{Massari}, {Koppelman}  \& {Helmi}}{{Massari}
  et~al.}{2019}]{2019A&A...630L...4M}
{Massari} D.,  {Koppelman} H.~H.,   {Helmi} A.,  2019, \mn@doi [\aap]
  {10.1051/0004-6361/201936135}, \href
  {https://ui.adsabs.harvard.edu/abs/2019A&A...630L...4M} {630, L4}

\bibitem[\protect\citeauthoryear{{Merritt}, {Piatek}, {Portegies Zwart}  \&
  {Hemsendorf}}{{Merritt} et~al.}{2004}]{2004ApJ...608L..25M}
{Merritt} D.,  {Piatek} S.,  {Portegies Zwart} S.,   {Hemsendorf} M.,  2004,
  \mn@doi [\apjl] {10.1086/422252}, \href
  {http://adsabs.harvard.edu/cgi-bin/nph-bib_query?bibcode=2004ApJ...608L..25M&db_key=AST}
  {608, L25}

\bibitem[\protect\citeauthoryear{{Milone} et~al.,}{{Milone}
  et~al.}{2016}]{2016MNRAS.455.3009M}
{Milone} A.~P.,  et~al., 2016, \mn@doi [\mnras] {10.1093/mnras/stv2415}, \href
  {https://ui.adsabs.harvard.edu/abs/2016MNRAS.455.3009M} {455, 3009}

\bibitem[\protect\citeauthoryear{{Miszalski}, {Shortridge}, {Saunders},
  {Parker}  \& {Croom}}{{Miszalski} et~al.}{2006}]{2006MNRAS.371.1537M}
{Miszalski} B.,  {Shortridge} K.,  {Saunders} W.,  {Parker} Q.~A.,   {Croom}
  S.~M.,  2006, \mn@doi [\mnras] {10.1111/j.1365-2966.2006.10777.x}, \href
  {https://ui.adsabs.harvard.edu/abs/2006MNRAS.371.1537M} {371, 1537}

\bibitem[\protect\citeauthoryear{{Moore}}{{Moore}}{1996}]{1996ApJ...461L..13M}
{Moore} B.,  1996, \mn@doi [\apjl] {10.1086/309998}, \href
  {http://adsabs.harvard.edu/abs/1996ApJ...461L..13M} {461, L13}

\bibitem[\protect\citeauthoryear{{Myeong}, {Vasiliev}, {Iorio}, {Evans}  \&
  {Belokurov}}{{Myeong} et~al.}{2019}]{2019MNRAS.488.1235M}
{Myeong} G.~C.,  {Vasiliev} E.,  {Iorio} G.,  {Evans} N.~W.,   {Belokurov} V.,
  2019, \mn@doi [\mnras] {10.1093/mnras/stz1770}, \href
  {https://ui.adsabs.harvard.edu/abs/2019MNRAS.488.1235M} {488, 1235}

\bibitem[\protect\citeauthoryear{{Nitadori} \& {Aarseth}}{{Nitadori} \&
  {Aarseth}}{2012}]{nitadoriaarseth2012}
{Nitadori} K.,  {Aarseth} S.~J.,  2012, \mn@doi [\mnras]
  {10.1111/j.1365-2966.2012.21227.x}, \href
  {http://esoads.eso.org/abs/2012MNRAS.424..545N} {424, 545}

\bibitem[\protect\citeauthoryear{{Odenkirchen} et~al.,}{{Odenkirchen}
  et~al.}{2001}]{2001ApJ...548L.165O}
{Odenkirchen} M.,  et~al., 2001, \mn@doi [\apjl] {10.1086/319095}, \href
  {https://ui.adsabs.harvard.edu/abs/2001ApJ...548L.165O} {548, L165}

\bibitem[\protect\citeauthoryear{{Palau} \& {Miralda-Escud{\'e}}}{{Palau} \&
  {Miralda-Escud{\'e}}}{2020}]{2020arXiv201014381P}
{Palau} C.~G.,  {Miralda-Escud{\'e}} J.,  2020, arXiv e-prints, \href
  {https://ui.adsabs.harvard.edu/abs/2020arXiv201014381P} {p. arXiv:2010.14381}

\bibitem[\protect\citeauthoryear{{Pe{\~n}arrubia}, {Varri}, {Breen}, {Ferguson}
   \& {S{\'a}nchez-Janssen}}{{Pe{\~n}arrubia}
  et~al.}{2017}]{2017MNRAS.471L..31P}
{Pe{\~n}arrubia} J.,  {Varri} A.~L.,  {Breen} P.~G.,  {Ferguson} A. M.~N.,
  {S{\'a}nchez-Janssen} R.,  2017, \mn@doi [\mnras] {10.1093/mnrasl/slx094},
  \href {https://ui.adsabs.harvard.edu/abs/2017MNRAS.471L..31P} {471, L31}

\bibitem[\protect\citeauthoryear{{Peebles}}{{Peebles}}{1984}]{1984ApJ...277..470P}
{Peebles} P.~J.~E.,  1984, \mn@doi [\apj] {10.1086/161714}, \href
  {https://ui.adsabs.harvard.edu/abs/1984ApJ...277..470P} {277, 470}

\bibitem[\protect\citeauthoryear{{Peuten}, {Zocchi}, {Gieles}, {Gualandris}  \&
  {H{\'e}nault-Brunet}}{{Peuten} et~al.}{2016}]{2016MNRAS.462.2333P}
{Peuten} M.,  {Zocchi} A.,  {Gieles} M.,  {Gualandris} A.,
  {H{\'e}nault-Brunet} V.,  2016, \mn@doi [\mnras] {10.1093/mnras/stw1726},
  \href {http://adsabs.harvard.edu/abs/2016MNRAS.462.2333P} {462, 2333}

\bibitem[\protect\citeauthoryear{Price-Whelan}{Price-Whelan}{2017}]{gala}
Price-Whelan A.~M.,  2017, \mn@doi [The Journal of Open Source Software]
  {10.21105/joss.00388}, 2

\bibitem[\protect\citeauthoryear{{Raghavan} et~al.,}{{Raghavan}
  et~al.}{2010}]{2010ApJS..190....1R}
{Raghavan} D.,  et~al., 2010, \mn@doi [\apjs] {10.1088/0067-0049/190/1/1},
  \href {https://ui.adsabs.harvard.edu/abs/2010ApJS..190....1R} {190, 1}

\bibitem[\protect\citeauthoryear{{Scarpa}, {Marconi}  \& {Gilmozzi}}{{Scarpa}
  et~al.}{2003}]{2003A&A...405L..15S}
{Scarpa} R.,  {Marconi} G.,   {Gilmozzi} R.,  2003, \mn@doi [\aap]
  {10.1051/0004-6361:20030762}, \href
  {https://ui.adsabs.harvard.edu/abs/2003A&A...405L..15S} {405, L15}

\bibitem[\protect\citeauthoryear{{Scarpa}, {Marconi}, {Gilmozzi}  \&
  {Carraro}}{{Scarpa} et~al.}{2007}]{2007A&A...462L...9S}
{Scarpa} R.,  {Marconi} G.,  {Gilmozzi} R.,   {Carraro} G.,  2007, \mn@doi
  [\aap] {10.1051/0004-6361:20066182}, \href
  {https://ui.adsabs.harvard.edu/abs/2007A&A...462L...9S} {462, L9}

\bibitem[\protect\citeauthoryear{{Simioni} et~al.,}{{Simioni}
  et~al.}{2018}]{2018MNRAS.476..271S}
{Simioni} M.,  et~al., 2018, \mn@doi [\mnras] {10.1093/mnras/sty177}, \href
  {https://ui.adsabs.harvard.edu/abs/2018MNRAS.476..271S} {476, 271}

\bibitem[\protect\citeauthoryear{{Sollima}, {Mart{\'\i}nez-Delgado},
  {Valls-Gabaud}  \& {Pe{\~n}arrubia}}{{Sollima}
  et~al.}{2011}]{2011ApJ...726...47S}
{Sollima} A.,  {Mart{\'\i}nez-Delgado} D.,  {Valls-Gabaud} D.,
  {Pe{\~n}arrubia} J.,  2011, \mn@doi [\apj] {10.1088/0004-637X/726/1/47},
  \href {https://ui.adsabs.harvard.edu/abs/2011ApJ...726...47S} {726, 47}

\bibitem[\protect\citeauthoryear{{Sollima}, {Baumgardt}  \& {Hilker}}{{Sollima}
  et~al.}{2019}]{2019MNRAS.485.1460S}
{Sollima} A.,  {Baumgardt} H.,   {Hilker} M.,  2019, \mn@doi [\mnras]
  {10.1093/mnras/stz505}, \href
  {https://ui.adsabs.harvard.edu/abs/2019MNRAS.485.1460S} {485, 1460}

\bibitem[\protect\citeauthoryear{{Tiongco}, {Vesperini}  \& {Varri}}{{Tiongco}
  et~al.}{2016}]{2016MNRAS.461..402T}
{Tiongco} M.~A.,  {Vesperini} E.,   {Varri} A.~L.,  2016, \mn@doi [\mnras]
  {10.1093/mnras/stw1341}, \href
  {http://adsabs.harvard.edu/abs/2016MNRAS.461..402T} {461, 402}

\bibitem[\protect\citeauthoryear{{Tiongco}, {Vesperini}  \& {Varri}}{{Tiongco}
  et~al.}{2018}]{2018MNRAS.475L..86T}
{Tiongco} M.~A.,  {Vesperini} E.,   {Varri} A.~L.,  2018, \mn@doi [\mnras]
  {10.1093/mnrasl/sly009}, \href
  {https://ui.adsabs.harvard.edu/abs/2018MNRAS.475L..86T} {475, L86}

\bibitem[\protect\citeauthoryear{{Trager}, {King}  \& {Djorgovski}}{{Trager}
  et~al.}{1995}]{1995AJ....109..218T}
{Trager} S.~C.,  {King} I.~R.,   {Djorgovski} S.,  1995, \mn@doi [\aj]
  {10.1086/117268}, \href
  {https://ui.adsabs.harvard.edu/abs/1995AJ....109..218T} {109, 218}

\bibitem[\protect\citeauthoryear{{Trenti}, {Padoan}  \& {Jimenez}}{{Trenti}
  et~al.}{2015}]{2015ApJ...808L..35T}
{Trenti} M.,  {Padoan} P.,   {Jimenez} R.,  2015, \mn@doi [\apjl]
  {10.1088/2041-8205/808/2/L35}, \href
  {https://ui.adsabs.harvard.edu/abs/2015ApJ...808L..35T} {808, L35}

\bibitem[\protect\citeauthoryear{{VandenBerg}, {Brogaard}, {Leaman}  \&
  {Casagrande}}{{VandenBerg} et~al.}{2013}]{VandenBerg13}
{VandenBerg} D.~A.,  {Brogaard} K.,  {Leaman} R.,   {Casagrande} L.,  2013,
  \mn@doi [\apj] {10.1088/0004-637X/775/2/134}, \href
  {http://adsabs.harvard.edu/abs/2013ApJ...775..134V} {775, 134}

\bibitem[\protect\citeauthoryear{{Vasiliev}}{{Vasiliev}}{2019a}]{2019MNRAS.484.2832V}
{Vasiliev} E.,  2019a, \mn@doi [\mnras] {10.1093/mnras/stz171}, \href
  {https://ui.adsabs.harvard.edu/abs/2019MNRAS.484.2832V} {484, 2832}

\bibitem[\protect\citeauthoryear{{Vasiliev}}{{Vasiliev}}{2019b}]{2019MNRAS.489..623V}
{Vasiliev} E.,  2019b, \mn@doi [\mnras] {10.1093/mnras/stz2100}, \href
  {https://ui.adsabs.harvard.edu/abs/2019MNRAS.489..623V} {489, 623}

\bibitem[\protect\citeauthoryear{{Wang}}{{Wang}}{2020}]{2020MNRAS.491.2413W}
{Wang} L.,  2020, \mn@doi [\mnras] {10.1093/mnras/stz3179}, \href
  {https://ui.adsabs.harvard.edu/abs/2020MNRAS.491.2413W} {491, 2413}

\bibitem[\protect\citeauthoryear{{Watkins}, {van der Marel}, {Bellini}  \&
  {Anderson}}{{Watkins} et~al.}{2015}]{2015ApJ...803...29W}
{Watkins} L.~L.,  {van der Marel} R.~P.,  {Bellini} A.,   {Anderson} J.,  2015,
  \mn@doi [\apj] {10.1088/0004-637X/803/1/29}, \href
  {https://ui.adsabs.harvard.edu/abs/2015ApJ...803...29W} {803, 29}

\bibitem[\protect\citeauthoryear{{de Boer}, {Gieles}, {Balbinot},
  {H{\'e}nault-Brunet}, {Sollima}, {Watkins}  \& {Claydon}}{{de Boer}
  et~al.}{2019}]{2019MNRAS.485.4906D}
{de Boer} T.~J.~L.,  {Gieles} M.,  {Balbinot} E.,  {H{\'e}nault-Brunet} V.,
  {Sollima} A.,  {Watkins} L.~L.,   {Claydon} I.,  2019, \mn@doi [\mnras]
  {10.1093/mnras/stz651}, \href
  {https://ui.adsabs.harvard.edu/abs/2019MNRAS.485.4906D} {485, 4906}

\bibitem[\protect\citeauthoryear{{van de Ven}, {van den Bosch}, {Verolme}  \&
  {de Zeeuw}}{{van de Ven} et~al.}{2006}]{2006A&A...445..513V}
{van de Ven} G.,  {van den Bosch} R.~C.~E.,  {Verolme} E.~K.,   {de Zeeuw}
  P.~T.,  2006, \mn@doi [\aap] {10.1051/0004-6361:20053061}, \href
  {http://adsabs.harvard.edu/abs/2006A%26A...445..513V} {445, 513}

\makeatother
\end{thebibliography}



\bsp	
\label{lastpage}
\end{document}